\documentclass[pra,onecolumn,nofootinbib,superscriptaddress]{revtex4-2}
\usepackage{calrsfs}
\usepackage{ulem}
\usepackage{graphics}
\usepackage{epstopdf}
\usepackage{bm}
\usepackage{bbm}
\usepackage{amssymb}
\usepackage{epstopdf}
\usepackage{graphicx}
\usepackage[caption=false]{subfig}
\usepackage{amsmath}
\usepackage{color}
\usepackage{array}
\usepackage{hyperref}

\DeclareMathOperator{\sinc}{sinc}

\begin{document}

\title{Attosecond delay metrology beyond the photon coherence time with spectrally resolved Hong-Ou-Mandel interferometry}

\author{Yingwen \surname{Zhang}}
\email{Corresponding author, yzhang6@uottawa.ca}
\affiliation{Nexus for Quantum Technologies, University of Ottawa, Ottawa ON, Canada, K1N6N5}
\affiliation{National Research Council of Canada, 100 Sussex Drive, Ottawa ON, Canada, K1A0R6}
\affiliation{Joint Centre for Extreme Photonics, National Research Council and University of Ottawa, Ottawa, Ontario, Canada}

\author{Kyle \surname{Jordan}}
\affiliation{Nexus for Quantum Technologies, University of Ottawa, Ottawa ON, Canada, K1N6N5}
\affiliation{National Research Council of Canada, 100 Sussex Drive, Ottawa ON, Canada, K1A0R6}
\affiliation{Joint Centre for Extreme Photonics, National Research Council and University of Ottawa, Ottawa, Ontario, Canada}

\author{Duncan \surname{England}}
\affiliation{National Research Council of Canada, 100 Sussex Drive, Ottawa ON, Canada, K1A0R6}
\affiliation{Joint Centre for Extreme Photonics, National Research Council and University of Ottawa, Ottawa, Ontario, Canada}

\author{Vincenzo \surname{Tamma}}
\affiliation{School of Mathematics and Physics, University of Portsmouth, Portsmouth PO1 3QL, United Kingdom}
\affiliation{Institute of Cosmology and Gravitation, University of Portsmouth, Portsmouth PO1 3FX, United Kingdom}

\author{Ebrahim \surname{Karimi}}
\affiliation{Nexus for Quantum Technologies, University of Ottawa, Ottawa ON, Canada, K1N6N5}
\affiliation{National Research Council of Canada, 100 Sussex Drive, Ottawa ON, Canada, K1A0R6}
\affiliation{Joint Centre for Extreme Photonics, National Research Council and University of Ottawa, Ottawa, Ontario, Canada}
\affiliation{Institute for Quantum Studies, Chapman University, Orange, California 92866, USA}

\author{Benjamin \surname{Sussman}}
\affiliation{National Research Council of Canada, 100 Sussex Drive, Ottawa ON, Canada, K1A0R6}
\affiliation{Nexus for Quantum Technologies, University of Ottawa, Ottawa ON, Canada, K1N6N5}
\affiliation{Joint Centre for Extreme Photonics, National Research Council and University of Ottawa, Ottawa, Ontario, Canada}

\begin{abstract} 
Hong–Ou–Mandel (HOM) interferometry enables delay estimation at the quantum precision limit but is traditionally constrained to path differences within the coherence time of the interfering photons. Here, we demonstrate single-shot path-delay sensing using spectrally resolved HOM interference, achieving a precision near the measurement Cramér–Rao bound over an operating range of approximately $\pm 300$\,$\mu$m, removing the conventional dynamic-range limitation imposed by the photon coherence window. By extracting delay information from the spectral interference fringes of spectrally entangled photon pairs, we retain near-optimal sensitivity over an operational range exceeding the photon coherence time by over two orders of magnitude. Using one million detected photon pairs, we achieve a time-delay precision of 13\,attosecond (4\,nm), while real-time operation (at 1\,Hz) with 2000 detected photons yields 330\,attosecond (100\,nm) precision. Because the estimator relies on fringe periodicity rather than absolute coincidence rates, the method is intrinsically robust to variations in interference visibility, eliminating the need for recalibration. As a practical demonstration, we measure the thickness of a nominally 300\,$\mu$m transmissive target with nanometer-scale precision. These results mark a significant step towards deploying quantum-enhanced HOM interferometry for high-precision displacement sensing in real-world applications.
\end{abstract}
\maketitle

%%%%%%%%%%%%%%%%%%%%%%%%%%  body  %%%%%%%%%%%%%%%%%%%%%%%%%%
\section{Introduction}

Precision optical delay measurement has long been a central task in optical metrology, underpinning applications in displacement sensing, thickness metrology, optical coherence tomography, frequency-comb ranging, and precision tests of fundamental physics~\cite{Bobroff_1993,Monnier2003,Berkovic2012,LIGO2016,Huang2025}. Classical first-order interferometric methods, including Michelson, Mach–Zehnder, white-light, and low-coherence interferometry, infer path-length differences from optical phase, fringe shifts, or coherence envelopes, and can achieve sub-nanometre resolution and high phase sensitivity. However, these approaches are fundamentally or practically constrained by phase ambiguity, environmental phase drift, mechanical instability, group delay dispersion, and classical intensity fluctuations. Quantum interferometric approaches offer complementary advantages by exploiting non-classical correlations and multi-photon interference~\cite{Giovannetti2011,Taylor2016,Pirandola2018}. These methods provide automatic dispersion cancellation~\cite{Abouraddy2002,Nasr2003,Teich2012}, intrinsic background rejection through coincidence detection~\cite{England2019,Zhang2020,GALLEGOTORROME2024}, which can exceed that provided by classical pulsed gating in the limit where fewer than one photon per pulse is generated~\cite{England2019,Drouin2026} and measurement precision governed by quantum-mechanical scaling laws rather than classical noise limits~\cite{Dowling2008,Afek2010,Kim2025}. These features have motivated the use of nonclassical light for high-precision delay estimation.

Quantum interference of indistinguishable and entangled photons lies at the heart of many fundamental phenomena in quantum mechanics and underpins a wide range of emerging quantum technologies. A paradigmatic example is Hong–Ou–Mandel (HOM) interference~\cite{Hong1987,Shih1988}, which arises from quantum interference between indistinguishable two-photon probability amplitudes at a beamsplitter (BS) and has become a cornerstone tool in photonic quantum information processing, quantum communication, and quantum metrology~\cite{Pan2012,Bouchard_2021,Jin2024,Jaouni2025}. Importantly, the interference does not fundamentally require the individual photon wave packets to overlap at the beamsplitter; postponed-compensation experiments have shown that HOM-type interference can be recovered even when the photons arrive at different times, provided that the distinguishing information between the alternative two-photon amplitudes is subsequently erased~\cite{Pittman1996}.

Beyond its foundational significance, HOM interferometry provides an exceptionally sensitive method for precision optical time-delay measurements. The characteristic signature of HOM interference is the suppression, or ``dip", in two-fold coincidence counts observed between the output ports of a balanced beamsplitter (BS) when the photons are rendered indistinguishable. The sharp variation of the coincidence rate near this dip forms the basis of conventional HOM-based delay metrology and has enabled attosecond-scale temporal delay measurement precision that can potentially approach the ultimate quantum precision limit~\cite{Dauler1999,Lyons2018,Sgobba2023}. The metrological properties of generalized HOM interferometry have also been investigated theoretically using time–frequency continuous variables. These studies have shown how spectral engineering of biphoton states can enhance the estimation of time and frequency shifts~\cite{Fabre2021,Meskine2024,Colin2025}, and have established symmetry-based conditions under which HOM measurements attain the quantum Fisher information~\cite{Descamps2023,Descamps2026}. The operating range of HOM-based sensing can also be extended through spectral engineering, with meter-scale, coherence-limited dynamic ranges demonstrated using MHz-bandwidth frequency-entangled photon pairs generated from resonant sources~\cite{Chen2026}. Despite these achievements, conventional HOM-based sensing suffers from a fundamental practical limitation: high sensitivity is confined to path-length differences within the coherence time of the interfering photons, i.e. the width of the HOM dip.

Several theoretical and experimental studies have shown that resolving additional photonic degrees of freedom,
such as momentum~\cite{Triggiani2024,Triggiani2025,Muratore2025,Guo2025,Wang2026,Maggio2026},~\cite{Gonzalez2026,Maggio2026_2}, frequency~\cite{Laibacher2018,Triggiani2023}, time~\cite{Maggio2025}, polarization~\cite{Maggio2025_2}, can recover interference signatures beyond the traditional HOM coherence (spatial or temporal) window. In particular, spectrally resolved two-photon measurements reveal structured correlation patterns that, in principle, encode time-delay information over a much wider range than is accessible from integrated coincidence counts alone~\cite{Gerrits2015,Jin2015,Orre2019,Zhang2021,Chen2023}, with recent work demonstrating more than a 20-fold enhancement in dynamic range with femtosecond time-delay precision using independent photon sources at telecom wavelengths and a time-of-flight fiber spectrometer~\cite{Brooks2026}. While these approaches have established the underlying metrological potential, achieving quantum-enhanced performance with spectrally resolved HOM interference in a practical, real-time measurement architecture remains an outstanding challenge. Previously, the fastest method to measure the joint spectral intensity (JSI) is through dispersive-fibre time-of-flight spectroscopy, however these still take in the order of minutes to measure the correlation pattern due to significant losses in the long fiber\cite{Brooks2026,Zielnicki2018} or dispersion engineered gratings\cite{Davis2017}.

In this work, we demonstrate single-shot path-delay sensing at the Cramér–Rao bound (CRB) using spectrally resolved HOM interferometry of spectrally entangled photons implemented with a custom-built two-photon spectrometer based on a time-tagging event camera~\cite{Zhang2021}. Instead of inferring the delay from the depth of the HOM dip, we extract it from the periodicity of the joint spectral interference fringes. In conventional spectrally integrated detection, the two-photon interference phase varies across the photon spectrum; once the path delay exceeds the photon coherence time, these frequency-dependent contributions oscillate rapidly across the spectrum, and upon spectral integration the interference term averages to zero, leaving a constant total coincidence rate thus restricting conventional HOM delay sensing to within the photon coherence window. Spectrally resolved detection avoids this cancellation by resolving the individual frequency components, revealing the underlying two-photon interference as spectral fringes that persist well beyond the photon coherence length (width of HOM dip), where the fringe oscillation frequency is directly proportional to the path delay. Importantly, because the delay is encoded in the spectral fringe periodicity, the measurement can be performed without scanning the interferometer through the HOM dip to calibrate its profile and interference visibility, as is typically required in conventional HOM-based delay sensing.

At a path delay of $\Delta z = 100$\,$\mu$m, we achieve a time-delay precision of 13\,as (4\,nm displacement) using approximately $10^6$ detected photon pairs, corresponding to 1.1 times the measurement Cramér–Rao bound of 11\,as. Real-time operation at a 1\,Hz update rate, using approximately 2000 detected photon pairs per measurement, yields a precision of 330\,as (100\,nm displacement). A measurement precision within 1.5 times the CRB, is maintained over an operating range from -300\,$\mu$m to 300\,$\mu$m, exceeding the 5.7\,$\mu$m wide HOM dip by over two orders of magnitude. Measurements remain possible over a range from -1000\,$\mu$m to 1000\,$\mu$m, approaching 350 times the HOM dip width, although the precision at the largest delays is reduced to approximately 15 times the CRB. In addition, because the delay is encoded in the spectral fringe periodicity rather than the absolute coincidence rate or fringe amplitude, the method requires only a one-time pixel-to-wavelength calibration of the spectrometer. Variations in interference visibility degrade the measurement precision without introducing a systematic bias in the inferred delay, thereby avoiding the need to rescan the HOM dip to characterize changes in the interference response. To the best of our knowledge, this also represents the first real-time implementation of spectrally resolved entangled-photon correlation measurements, enabled by the TPX3CAM event camera and the integration of photon-event reconstruction and coincidence-correlation processing, developed in \cite{Zhang2021}, into the camera-control software, allowing both data acquisition and processing to be performed in real time. As a practical demonstration, we perform thickness metrology of a 300\,$\mu$m transmissive target with nanometer-scale precision in a single shot. These results establish spectrally resolved HOM interferometry as a practical platform for large dynamic range, high-precision quantum displacement sensing and represent a significant step towards its practical deployment in real-world applications.

\section{Method}
\subsection{Experimental Concept} 
\begin{figure}
    \centering
    \includegraphics[width=1\linewidth]{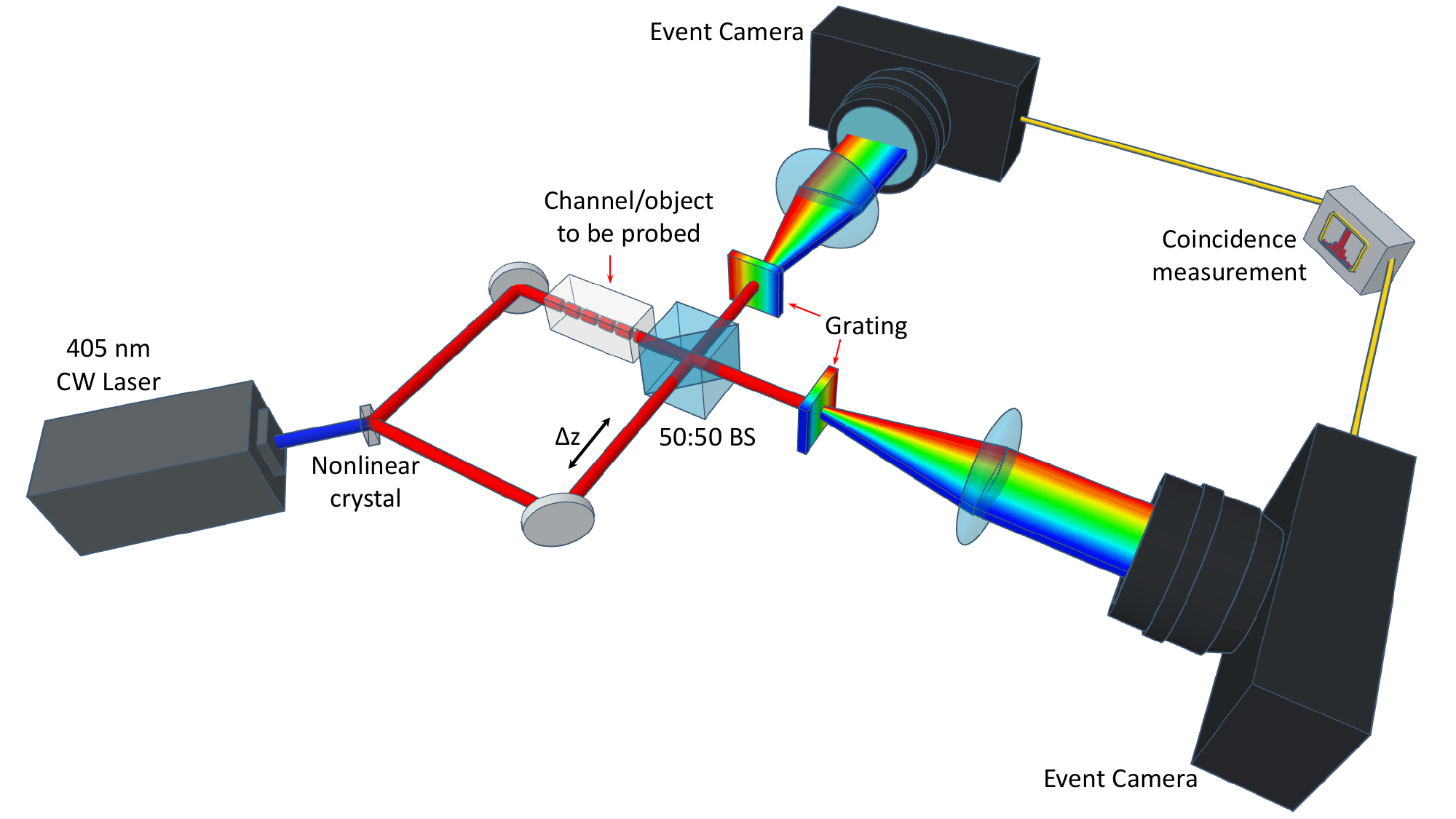}
    \caption{{\bf Conceptual setup for spectrally resolved HOM interferometry.} Broadband spectrally entangled photon pairs are generated in a nonlinear crystal pumped by a 405\,nm CW laser. One photon propagates through a channel or object that introduces an optical path delay to be monitored or measured. The two photons are then recombined at a 50:50 beamsplitter (BS), where they undergo HOM interference. The output photons are directed to custom spectrometers based on time-tagging event cameras. In each spectrometer, the photons are spectrally dispersed by a diffraction grating and imaged onto the camera. SPDC photon pairs are identified through coincidence analysis using their detection times on the cameras.}
    \label{Fig1}
\end{figure}

The conceptual setup for spectrally resolved HOM interferometry is shown in Fig.\,\ref{Fig1}. A 405\,nm continuous-wave (CW) laser pumps a 1\,mm-thick Type-0 periodically poled potassium titanyl phosphate (ppKTP) crystal to generate broadband spectrally entangled photon pairs via spontaneous parametric down-conversion (SPDC) with the two-photon state given by
\begin{equation}
    |\psi\rangle = \int d\omega_1 d\omega_2 \Phi(\omega_1,\omega_2) \hat{a}^\dagger_1(\omega_1)\hat{a}^\dagger_2(\omega_2)|\text{vac}\rangle,
\end{equation}
where $\omega_1$ and $\omega_2$ denotes the frequencies of the two photons, $\hat{a}_j^\dagger(\omega)$ is the creation operator for a photon of frequency $\omega$ in path $j$, and $|\mathrm{vac}\rangle$ denotes the vacuum state. The joint spectral amplitude $\Phi(\omega_1,\omega_2)$ encodes the spectral correlations of the photon pair and is determined by the pump-envelope function and the phase-matching response of the nonlinear crystal. Its squared modulus, $\left|\Phi(\omega_1,\omega_2)\right|^2$, defines the JSI, which gives the probability density for detecting one photon at frequency $\omega_1$ and its partner at frequency $\omega_2$. For a source that is symmetric under exchange of the two photons, the joint spectral amplitude satisfies $\Phi(\omega_1,\omega_2)=\Phi(\omega_2,\omega_1)$.

The photon pairs are then spatially separated with one photon of the pair propagating through a channel or an object whose optical path delay $\Delta t$ is to be monitored or measured. The photon pairs are then recombined at a 50:50 BS to undergo HOM interference. The probability distributions for spectrally entangled photon pairs exiting the HOM interferometer in the bunched (same exit port) and antibunched (different exit ports) configurations, are given by~\cite{Ou1988,Chen2019}
\begin{align}
    P_B(\Delta t,\omega_1,\omega_2) &= \frac{1}{2}\left|\Phi(\omega_1,\omega_2)\right|^2\bigl[1 + V(\Delta t)\cos[(\omega_1-\omega_2)\Delta t]\bigr] \quad \text{bunching}\nonumber\\
    P_A(\Delta t,\omega_1,\omega_2) &= \frac{1}{2}\left|\Phi(\omega_1,\omega_2)\right|^2\bigl[1 - V(\Delta t)\cos[(\omega_1-\omega_2)\Delta t]\bigr] \quad \text{antibunching}.
    \label{Cprob}
\end{align}
Here, $V(\Delta t) = \eta^2\xi(\Delta t)$ is the effective HOM interference visibility, where $\eta = \eta_{\text{BS}}\eta_{\text{ind}}\eta_{\text{acc}}$. Specifically, $\eta_{\text{BS}} = \sqrt{2TR/(T^2+R^2)}$ accounts for the non-ideal beam-splitter splitting ratio $T:R$, while $\eta_{\text{ind}}$  describes the residual photon distinguishability arising from factors such as polarization mismatch and imperfect spatial-mode overlap; and $\eta_{\text{acc}}$  accounts for the reduction in visibility due to accidental coincidences. The factor $0\leq\xi(\Delta t)\leq1$ accounts for the finite spectral resolution of the spectrometer, which reduces the resolvability of spectral interference fringes with oscillation frequency set by $\Delta t$. For notational simplicity, the effective visibility $V$ is henceforth understood to be a function of $\Delta t$, and its explicit delay dependence will be omitted unless otherwise stated.

If one assumes a Gaussian spectrometer point-spread function with root-mean-square (RMS) spectral width $\sigma_{\mathrm{PSF}}$ and perfectly energy-anticorrelated photon pairs such that $\omega_1+\omega_2 = \omega_p$, with $\omega_p$ the pump photon frequency, we obtain (see the Supplementary Material for detailed derivation)
\begin{equation}
    \xi(\Delta t) = \exp(-2\sigma_{\mathrm{PSF}}^2\Delta t^2).
    \label{SPSF}
\end{equation}
One can see the probability distributions given by Eq.\eqref{Cprob} exhibit oscillatory interference fringes whose frequency is proportional to $\Delta t$. Importantly, these spectral interference fringes persist well beyond the coherence time of the SPDC photons, enabling delay measurement over a significantly extended range.

\begin{figure}
    \centering
    \includegraphics[width=1\linewidth]{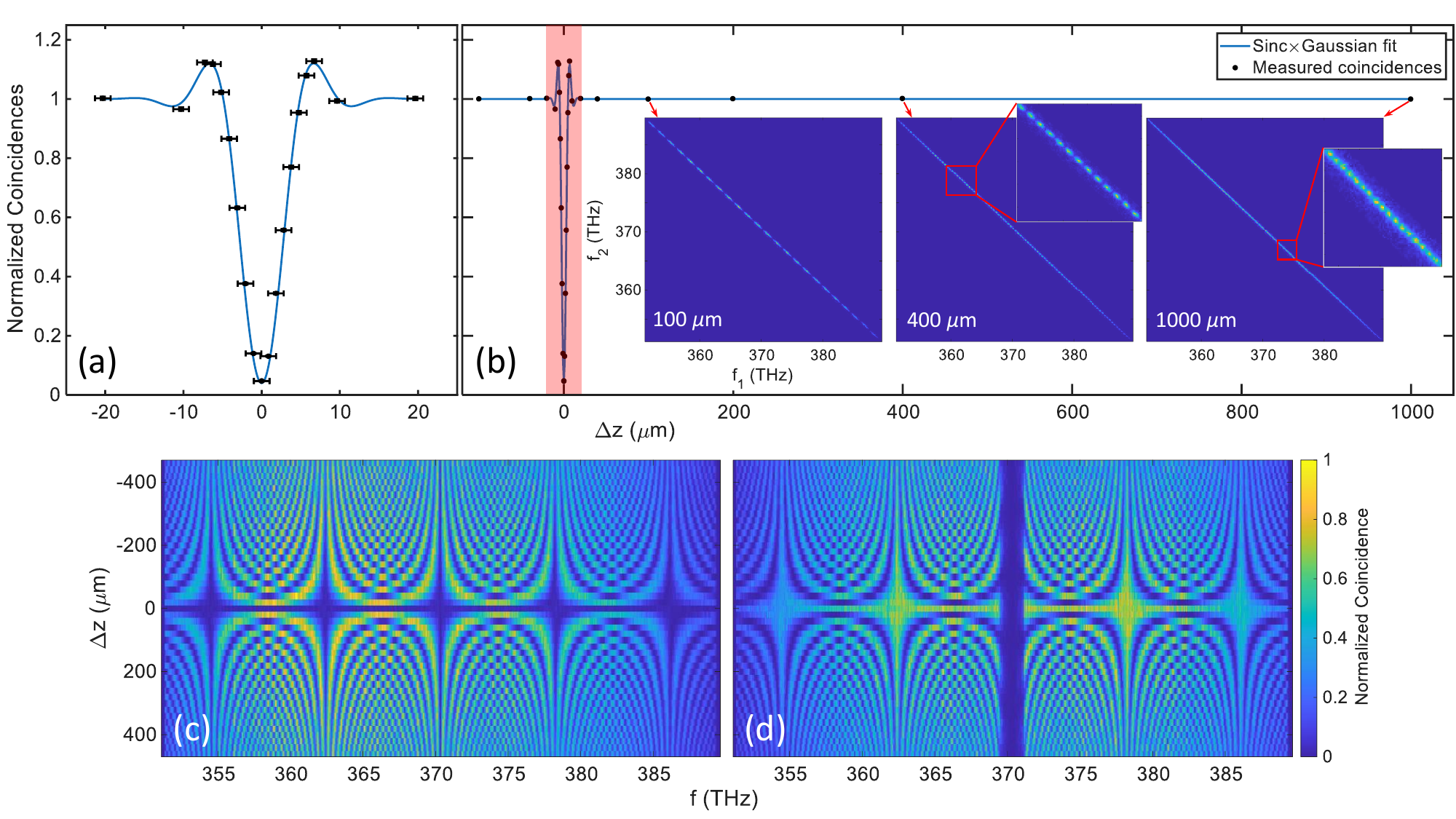}
    \caption{{\bf JSI of photon pairs exiting the HOM interferometer.} (a) Spectrally unresolved HOM interference dip measured without the use of spectral filters through avalanche photon diode detectors giving a HOM interference visibility $V = 0.92$. Horizontal error bar indicates the quoted actuator accuracy of 1\,$\mu$m used to adjust the path delay. Blue line is the fitted function $1-\alpha\sinc(\beta\Delta z)\exp(-\Delta z^2/\gamma)$, giving a 5.7\,$\mu$m FWHM of the dip. (b) HOM dip measured over a much longer range with the highlighted red region showing the range displayed in (a). Inset shows the measured JSI at $\Delta z = 100$\,$\mu$m, $400$\,$\mu$m and $1000$\,$\mu$m. The RMS width of the correlation is 0.11\,THz (3.2 pixels) over a 36.5\,THz (1024 pixel) range (c, d) Cross-sections of the measured JSI projected onto $f_1$ plotted as a function of path delay $\Delta z$ for antibunching (c) and bunching (d).  The dark vertical band in (d) is due to the camera being unable to resolve two photons detected in close proximity to each other~\cite{Zhang2021,Vidyapin2022}, it has a negligible effect on the accuracy of the inferred delay, as discussed in more detail in the Supplementary Material.}
    \label{Fig2}
\end{figure}

To observe the spectral interference fringes, photons from the two output ports of the beamsplitter are directed to spectrally resolving detectors, depicted conceptually in Fig.~\ref{Fig1} as two separate time-tagging event-camera spectrometers. The event camera records the spatial position and arrival time of individual photon detections with nanosecond timing resolution. Coincidence analysis on the photon's time of detection by the cameras allows the joint spectral correlations of photon pairs to be reconstructed as shown in Fig.~\ref{Fig2}. In the actual experimental implementation, rather than using two separate cameras, both beamsplitter outputs were directed to a single event camera, with the two photon spectra mapped onto spatially separated regions of the sensor. Temporal and spectral correlations were then measured between photon detections in these two regions. Compared with our previous time-tagging spectrometer implementation~\cite{Zhang2021}, the present system incorporates a more efficient photon event processing software, enabling real-time spectral-correlation processing and delay estimation as well as pixel super-resolution through a clustering and centroiding algorithm resulting in approximately fourfold higher spectral resolution (see the Supplementary Material for a detailed experimental setup).

For this experiment, degenerate, frequency-anticorrelated photon pairs centered around $810$\,nm with $19.76\pm0.04$\,nm RMS spectral bandwidth is generated. Figure~\ref{Fig2}(a) shows the spectrally unresolved HOM interference dip measured through avalanche photon diode detectors without the use of any spectral filters without the use of any spectral filters giving $V=0.920\pm0.007$. The shape of the HOM dip is governed by the Fourier transform of the photon-pair frequency-difference distribution. Approximating this distribution as a rectangular spectral profile convolved with a Gaussian yields a sinc-shaped interference envelope modulated by a Gaussian function. Fitting the function $1-\alpha\sinc(\beta\Delta z)\exp(-\Delta z^2/\gamma)$ ($\alpha = 0.96$, $\beta = 0.204$\,$\mu$m$^{-1}$, $\gamma =85$\,$\mu$m$^{-2}$) to the measured dip gives a FWHM width of $5.7\pm1.2$\,$\mu$m. Figure~\ref{Fig2}(b) shows the measured JSI at distances much larger than the dip width at $\Delta z = 100$\,$\mu$m, $400$\,$\mu$m and $1000$\,$\mu$m in which the spectral interference fringes can be clearly observed.

The achievable delay range is ultimately constrained at large delays by the spectral resolution of the spectrometer which has a spectral resolution of $\sim$0.26\, nm over the $(772.08–852.06)\pm0.02$\,nm spectral range, this corresponds to a characteristic resolution-limited delay of $|\Delta z| \approx 1.15$\,mm (See Supplementary Material for details). In practice, with $N_c = 10^4$ detected photon pairs, the oscillation frequency can be reliably extracted using a fast Fourier transform (FFT) upto $|\Delta z| \approx 1$\,mm, while with $N_c = 10^5$, the FFT peak remains resolvable up to approximately 1.4\,mm. Beyond this range, the progressively decreasing fringe contrast causes the FFT peak to become indistinguishable from the noise. On the other hand at small delays, the FFT estimator becomes progressively less well localized when fewer than approximately one complete spectral oscillation is contained within the measured bandwidth, corresponding to $|\Delta z| \lesssim 4.1$\,$\mu$m for the present system. This lower value represents a practical crossover rather than a fundamental cutoff, since it overlaps with the conventional HOM-dip region (5.7\,$\mu$m FWHM), where the spectrally integrated coincidence rate can instead be used for delay estimation.

\subsection{Quantum Precision Limit}

The Fisher information associated with frequency-resolved measurements is given by~\cite{Chen2019,Chen2023,Triggiani2023}
\begin{align}
        \mathcal{F}(\Delta t) &= \int\int d\omega_1d\omega_2 \left\{\frac{1}{P_B(\Delta t,\omega_1,\omega_2)}\left[\frac{d}{d\Delta t} P_B(\Delta t,\omega_1,\omega_2)\right]^2\right. \nonumber\\
        &\quad + \left.\frac{1}{P_A(\Delta t,\omega_1,\omega_2)}\left[\frac{d}{d\Delta t} P_A(\Delta t,\omega_1,\omega_2)\right]^2\right\}.
\end{align}
Substituting in Eq.\eqref{Cprob},while accounting for the delay dependence of the visibility, gives
\begin{equation}
    \mathcal{F}(\Delta t) = \int\int d\omega_1d\omega_2 \left|\Phi(\omega_1,\omega_2)\right|^2 \frac{\left[V'\cos(\Delta\omega\Delta t) - V\Delta\omega\sin(\Delta\omega\Delta t)\right]^2}{1-V^2\cos^2(\Delta\omega\Delta t)},
    \end{equation}
where $\Delta\omega = \omega_1 - \omega_2$ and $V' = dV/d\Delta t$. For the Gaussian spectrometer response introduced in Eq.\eqref{SPSF} the Fisher information becomes
\begin{equation}
    \mathcal{F}(\Delta t) = \int\int d\omega_1d\omega_2 \left|\Phi(\omega_1,\omega_2)\right|^2 \frac{V^2\left[4\sigma_{\mathrm{PSF}}^2\Delta t \cos(\Delta\omega\Delta t) + \Delta\omega\sin(\Delta\omega\Delta t)\right]^2}{1-V^2\cos^2(\Delta\omega\Delta t)}.
    \label{Fish1}
\end{equation}
In the large-delay regime, the spectral fringes oscillate rapidly across the bandwidth. Under this approximation and by assuming perfect spectral anticorrelation, the Fisher information simplifies to (see the Supplementary Material for detailed derivation)
\begin{align}
    \mathcal{F}(\Delta t) &\approx \left(1-\sqrt{1-V^2}\right)(4\sigma^2 + \Omega^2) \nonumber\\
        &= \left(1-\sqrt{1-\eta^4 \exp(-4\sigma_{\mathrm{PSF}}^2\Delta t^2)}\right)(4\sigma^2 + \Omega^2)
    \label{Fish2}
\end{align}
where $\sigma = \sqrt{\langle\omega^2\rangle-\langle\omega\rangle^2}$ is the RMS spectral bandwidth of the SPDC photons and $\Omega = |\langle\omega_1\rangle - \langle\omega_2\rangle|$ is the difference between the central frequencies of the two photons. In the present experiment, degenerate photon pairs are used, so $\Omega=0$ and the Fisher information is determined solely by the SPDC bandwidth.

In the ideal case where $V=1$, the Fisher information becomes
\begin{equation}
    \mathcal{F}_0 = 4\sigma^2 + \Omega^2. 
    \label{Fish3}
\end{equation}
Under the ideal conditions assumed in Eq.\eqref{Fish3}, the measurement Fisher information $\mathcal{F}_0$ is equal to the quantum Fisher information $\mathcal{H}$ of the delay-encoded two-photon state, which is determined by\cite{Triggiani2023, Chen2019}.
\begin{equation}
    \mathcal{H} = 4\left[\left\langle\left.\frac{\partial\psi}{\partial\Delta t}\right|\frac{\partial\psi}{\partial\Delta t}\right\rangle - \left|\left\langle\psi\left|\frac{\partial\psi}{\partial\Delta t}\right.\right\rangle\right|^2\right] = 4\sigma^2 + \Omega^2 = \mathcal{F}_0. 
    \label{Fish4}
\end{equation}

Equations~\eqref{Fish2}-\eqref{Fish4} shows that the delay-estimation sensitivity is determined by two distinct spectral resources. The first contribution, $4\sigma^2$, arises from the photon bandwidth and reflects the fact a broader spectrum spans a larger range of difference frequencies $\omega_1-\omega_2$, so that for a given delay the interference phase covers a larger range and more spectral-fringe oscillations are observed. These larger difference-frequency components also accumulate phase more rapidly with changes in delay, thereby increasing the available timing information. The second contribution, $\Omega^2$, arises from nondegeneracy between the signal and idler photons. A larger separation between their central frequencies introduces an additional biphoton beat note, which further enhances the sensitivity even for a fixed individual-photon bandwidth. Thus, both increasing the photon bandwidth and using nondegenerate photon pairs provide routes to improving the ultimate delay precision. Importantly, Eq.\eqref{Fish2}  shows that, when the effects of finite spectrometer resolution $\sigma_{\mathrm{PSF}}$ can be neglected, the effective interference visibility becomes independent of $\Delta t$, and the Fisher information approaches a constant value in the large-delay regime. This indicates that spectrally resolved HOM measurements can, in principle, maintain the same delay sensitivity over a range far exceeding the conventional HOM coherence window. 

The corresponding precision limit for estimating the time delay is~\cite{Chen2019,Chen2023} 
\begin{equation}
    \mathrm{SD}(\Delta t) \geq \frac{1}{\sqrt{\mathcal{F}N_c}}\geq\frac{1}{\sqrt{\mathcal{H}N_c}},
    \label{CRB}
\end{equation}
where $N_c$ is the number of photon pairs. The first inequality gives the measurement CRB, while the second corresponds to the quantum CRB and the two coincide in the ideal limit $\mathcal{F} = \mathcal{F}_0 = \mathcal{H}$.

For finite spectral anticorrelation, the variance $\langle(\omega_1-\omega_2)^2\rangle$ is reduced from the ideal value of $4\sigma^2+\Omega^2$, leading to a corresponding reduction in the available Fisher information. As the spectral anticorrelation becomes weaker, the result approaches the uncorrelated limit $\langle(\omega_1-\omega_2)^2\rangle = 2\sigma^2+\Omega^2$~\cite{Triggiani2023}, corresponding to a separable joint spectral distribution $S(\omega_1)S(\omega_2)$. This limit can also be realized with classical broadband light, such as phase-averaged coherent or thermal light, which can produce analogous frequency-resolved second-order interference patterns. In addition, for coherent light fringe visibility is bounded by $V\leq 1/2$~\cite{Lo2012,Kim2020_2} and for thermal light the bound is further reduced to $V\leq 1/3$~\cite{Tamma2016,Cassano2017,Ihn2017,Lee2020}. In the case of degenerate photons ($\Omega=0$), this corresponds to a 15 times reduction to the Fisher information when using coherent light and 35 times reduction when using thermal light. For the present experiment, with a measured visibility of $V(0) = 0.92$, the corresponding enhancements are approximately $9.1\times$ and $21\times$ respectively.

Although classical sources can provide much higher photon fluxes, our SPDC source already operates close to the practical saturation limit of the TPX3CAM camera. Since coherent classical light would require approximately 15 times more detected photons to reach the same CRB-limited precision, this would exceed the count-rate capability of the present detector. Thus, in the current implementation, the brightness advantage of classical illumination cannot be fully exploited.

In this work, we determine the delay $\Delta t$ by applying a FFT to the measured spectral interference pattern, rather than performing a maximum-likelihood estimation (MLE). In this approach, the fringe oscillation frequency extracted from the FFT directly determines the time delay, without requiring an accurate model of the full measured probability distribution as in MLE. This makes the FFT based estimator robust to experimental imperfections that affect the amplitude envelope and interference visibility, but are not fully captured by the probability model. To localize the corresponding frequency peak with high numerical precision, we evaluate the Fourier spectrum on a finely sampled grid of 65536 points by zero padding the measured data. While zero padding does not add information or reduce the fundamental statistical uncertainty, it provides an efficient interpolation of the discrete Fourier transform and enables sub-bin localization of the spectral peak. For independently measured data well described by a single-frequency cosine/sine modulation, a finely sampled FFT peak estimator can approach the performance of MLE~\cite{Rife1974}, while avoiding the convergence issues that can arise in iterative maximum-likelihood fitting when the probability-distribution model is imperfect.

\section{Results}
\begin{figure}
    \centering
    \includegraphics[width=1\linewidth]{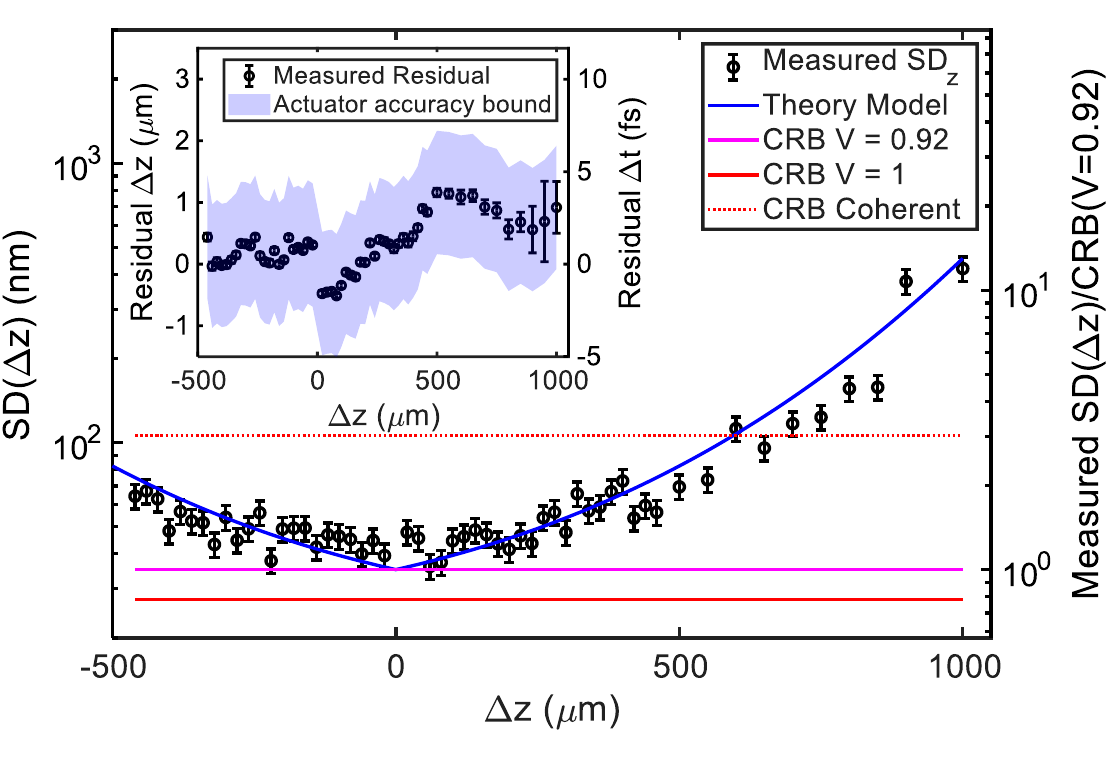}
    \caption{{\bf Measurement precision of spectrally resolved HOM interferometry.} Standard deviation of the estimated delay as a function of interferometer path difference $\Delta z$, obtained from $N_c = 10^4$ detected photon pairs with 50 repeated measurements at each delay. The left axis shows the displacement precision $\text{SD}(\Delta z)$ while the right axis shows ratio of the measured $\text{SD}(\Delta z)$ and the CRB at $V=0.92$, the highest measured for the system. Error bars represent the statistical uncertainty in the measured standard deviation due to the finite number of repeats. The purple line indicates the CRB calculated using the highest experimentally measured visibility $V=0.92$ at $\Delta z=0$ with avalanche photo-diode detectors, the red line shows the ideal CRB for $V=1$ and the dashed red line gives the CRB for a broadband classical coherent state. Blue curve is the theoretical model from using Eq.~\eqref{Fish2}. The inset shows the residual error between measured and expected $\Delta z$ with the blue shaded region indicates the specified positioning accuracy of the motorized translation stage (1\,$\mu$m). Note that the SPDC photons have a coherence length of 5.7\,$\mu$m, while the measurement on $\Delta z$ can be made for length up to 1\,mm.}
    \label{Fig3}
\end{figure}

We first evaluate the precision of displacement estimation using spectrally resolved HOM interferometry. Figure~\ref{Fig3} shows the standard deviation of the measured displacement $\mathrm{SD}(\Delta z) = c\mathrm{SD}(\Delta t)$ (with $c$ the speed of light), as a function of the interferometer path difference $\Delta z$, obtained from $N_c = 10^4$ detected bunched and antibunched photon pairs with 50 repeated measurements at each delay. The theoretical modeled CRB of the spectrally correlated two-photon state is calculated using Eq.\,\eqref{Fish2} and \eqref{CRB}, with the highest experimentally measured visibility $V(0) = 0.92$ ($\eta=0.96$) using avalanche photo-diode detectors. Given the beamsplitter has a quoted splitting ratio of T:R = 57:43 at 810\,nm, the theoretical upper limit of $V(0)$ for this setup is 0.96 ($\eta=0.98$). The remaining visibility reduction is expected to be mainly from residual spatial-mode mismatch associated with fiber coupling, as well as small contributions from polarization and spectral mismatch between the two photons. The inset of Fig.~\ref{Fig3} shows the residual difference between the measured and expected $\Delta z$, confirming the accuracy of the measurement across the tested delay range. Notably, the achieved precision remains below $1.5$ times the two photon CRB at $V(0)=0.92$ up to $\sim300$\,$\mu$m and outperforms the CRB limit for a coherent state up to $\sim600$\,$\mu$m. Delay measurement can still be performed up to 1\,mm, although with precision dropping to around $15$ times the CRB. These path differences far exceeds the HOM dip width of 5.7\,$\mu$m (Fig.\,\ref{Fig2}(a)), demonstrating the extended dynamic range enabled by spectrally resolved HOM measurements. 

The reduced precision at large displacements arises from the increasing density of spectral interference fringes with increasing $\Delta z$, which eventually approaches the Nyquist spectral-resolution limit of the camera. This finite-resolution effect is captured by the factor $\xi$ introduced previously, which quantifies the resolvability of the spectral interference fringes. As $\Delta z$ increases, $\xi$ decreases, leading to a loss of resolvable fringe contrast and a corresponding reduction in the available Fisher information, analogous to the reduction in measurement precision caused by decreased HOM interference visibility $V=\eta^2\xi$. The theoretical model of this decline, is shown by the blue curve in Fig.\,\ref{Fig3}, with details given in the Supplementary Material.

\begin{figure}
    \centering
    \includegraphics[width=1\linewidth]{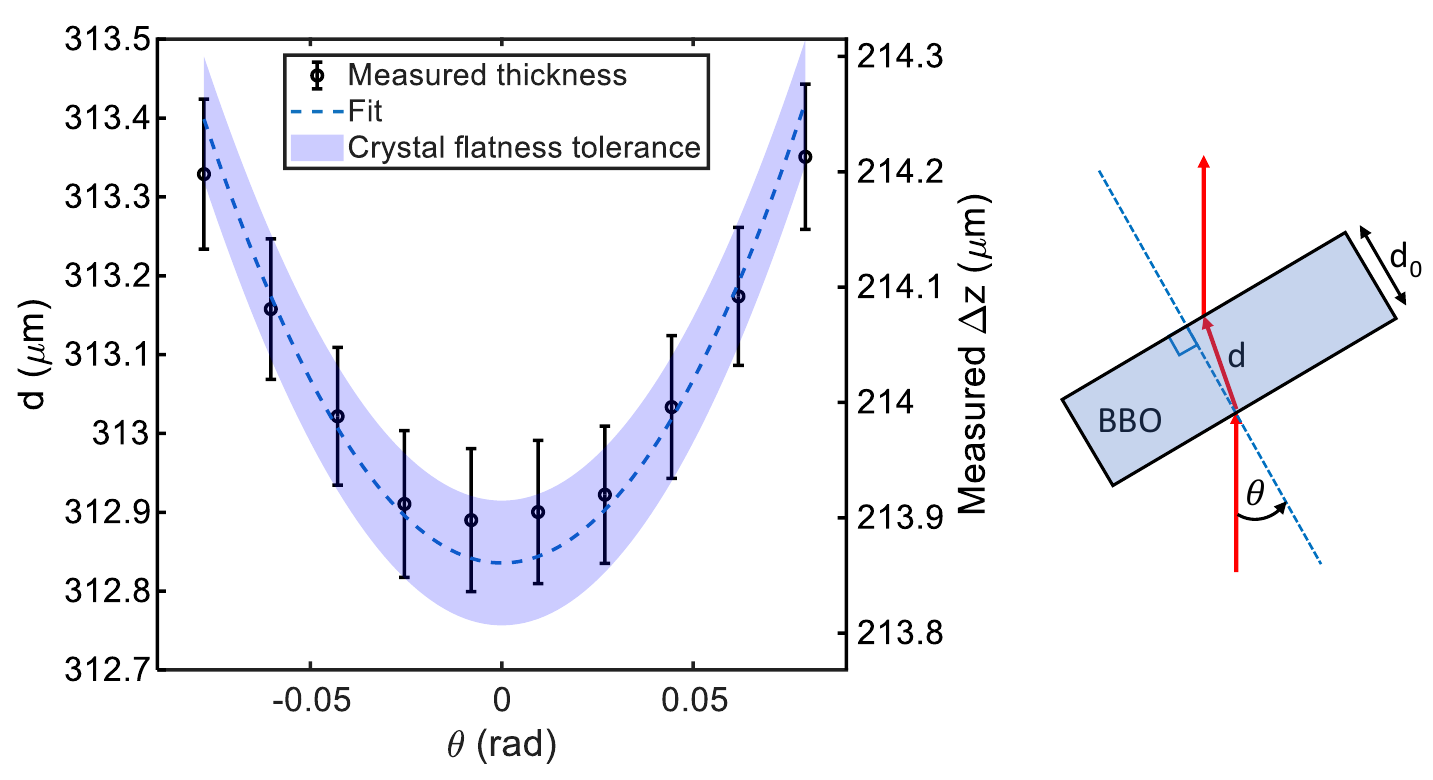}
    \caption{{\bf Thickness measurement of a BBO crystal.} Measured thickness $d$ of a BBO crystal (left axis) and the corresponding interferometer path difference $\Delta z$ (right axis) as a function of the rotation angle $\theta$. Error bars are obtained from 60 repeated measurements using $2.5\times10^4$ detected photons pair per measurement. The dashed blue curve shows a fit to the data using Eq.\,\eqref{BBOz}, yielding a zero-angle thickness of $d_0 = 312.85\pm0.03$\,$\mu$m. The blue shaded region indicates the manufacturer-specified surface flatness tolerance of the crystal ($\lambda/8$ at 633\,nm). The schematic on the right illustrates the geometry used to relate the measured optical delay to the crystal thickness. }
    \label{Fig4}
\end{figure}

To demonstrate a practical metrology application of the technique, we measured the thickness of a BBO crystal using the optical delay introduced by inserting the crystal into one arm of the interferometer. By tracking the change in the path delay as the crystal is rotated relative to the beam axis, under small rotation angle approximation, the measured path difference is expected to follow the geometric model (see Supplementary Material for the derivation)
\begin{equation}
    \Delta z(\theta) = \frac{d_0}{\cos\left(\theta n_A/n_B\right)}\left[n_B-n_A\cos\left(\frac{\theta(n_B-n_A)}{n_B}\right)\right], 
    \label{BBOz}
\end{equation}
from which the crystal thickness $d_0$ can be determined. Here $n_A$ and $n_B$ are the refractive indices of air and BBO, respectively and $\theta$ is the rotation angle of the BBO crystal about the rotation-stage axis lying in the plane of the crystal surface perpendicular to the plane of incidence. The zero position, $\theta=0$, corresponds to normal incidence, where the crystal surface normal is aligned with the incident beam. Positive and negative $\theta$ correspond to rotations to either side of this position according to the rotation-stage convention. Since the expression is symmetric with respect to $\theta \rightarrow -\theta$, the choice of sign does not affect the calculated delay.

Figure~\ref{Fig4} shows the measured crystal thickness $d$ as a function of the crystal rotation angle $\theta$. Prior to the measurements, the interferometer is set to a reference displacement of $\Delta z_{\mathrm{ref}} = -48.08\pm0.03$\,$\mu$m, against which the subsequent delays from the BBO are measured. Each data point is obtained from 60 repeated measurements using $2.5 \times 10^4$ detected photon pairs. Fitting Eq.\eqref{BBOz} to the measured delay using $n_A = 1.00028$, the refractive index of air at 810\,nm, and $n_B = 1.6839$, the ordinary group refractive index of BBO at 810\,nm calculated from the Sellmeier relation of~\cite{Zhang2000}, yields a zero-angle thickness of $d_0 = 312.85\pm0.03$\,$\mu$m. The slight deviations of the measured $\Delta z(\theta)$ from the fit of Eq.\eqref{BBOz} may arise from the beam probing different regions of the crystal surface as the rotation angle $\theta$ changes. The manufacturer specifies a surface flatness of $\lambda/8$ at 633\,nm, with the corresponding tolerance indicated by the blue shaded region in Fig.~\ref{Fig4}.

As an independent experimental check, classical spectrally resolved interferometry was performed using a femtosecond laser at central wavelengths ranging from 774 to 845\,nm, yielding a thickness of $313\pm2$\,$\mu$m, where the uncertainty reflects the observed wavelength-dependent variations. A separate mechanical micrometer measurement gave an approximate thickness of $315\pm3$\,$\mu$m (this is an estimated uncertainty since the manufacturer's calibration accuracy is unavailable). Both measurements are consistent with the thickness obtained from spectrally resolved HOM interferometry. These results also fall within the manufacturer's specified nominal thickness of 300\,$\mu$m with a tolerance of $\pm10\%$.

It should be noted that the inferred crystal thickness depends on the adopted Sellmeier relation. Using an alternative relation from~\cite{Eimerl1987}, which gives $n_B = 1.6850$ at 810\,nm, yields $d_0 = 312.35\pm0.03$\,$\mu$m. This model dependence should therefore be considered when applying the technique to measurements requiring accurate determination of absolute thickness. Further details of the independent measurements are provided in the Supplementary Material.

\begin{figure}
    \centering
    \includegraphics[width=1\linewidth]{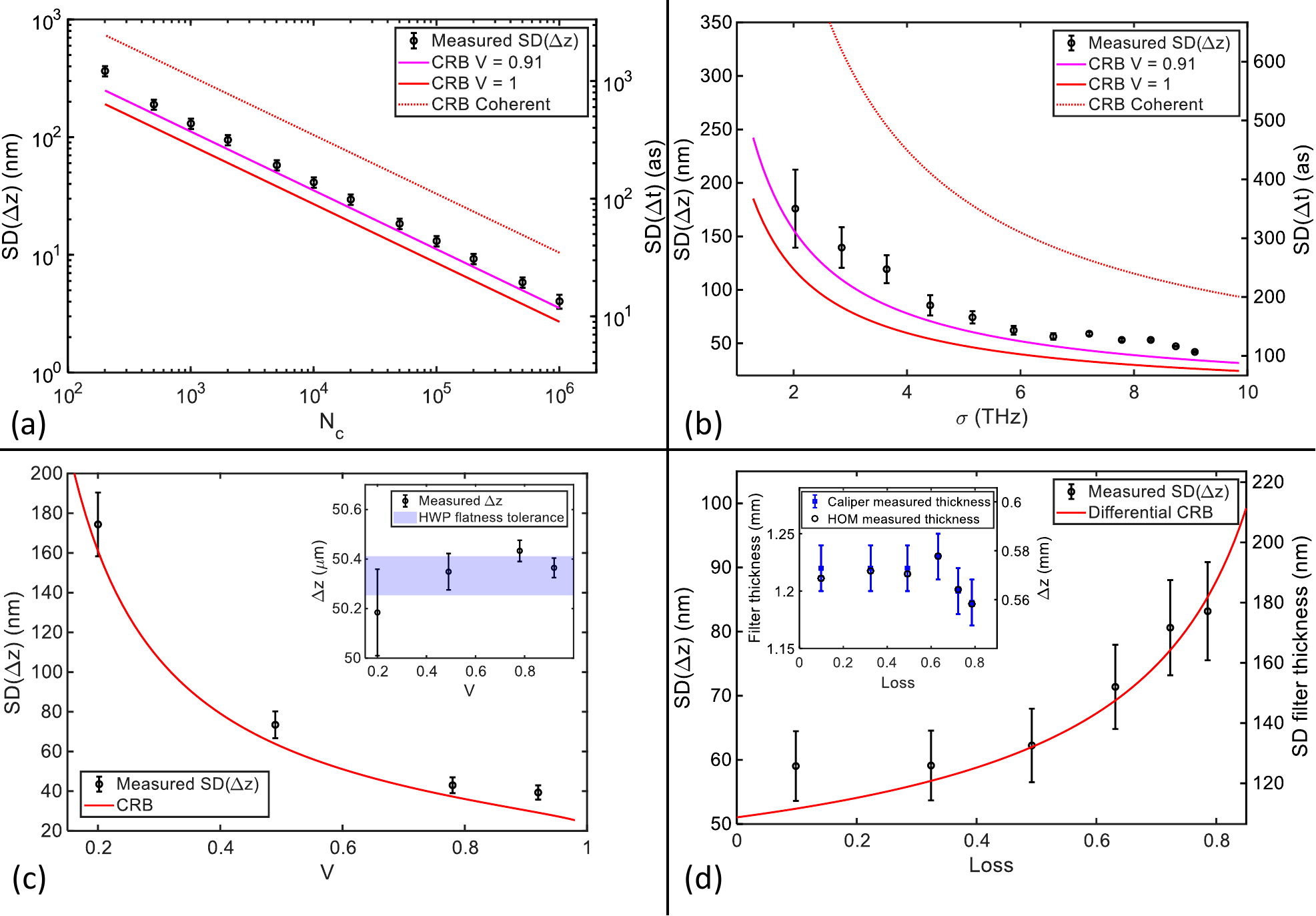}
    \caption{{\bf Measurement performance under varying experimental parameters.} (a) Standard deviation of the measured displacement $\mathrm{SD}(\Delta z)$ as a function of the number of detected coincidences $N_c$ at $\Delta z = 100$\,$\mu$m with a measured $V=0.91$. (b) $\mathrm{SD}(\Delta z)$ as a function of the RMS bandwidth $\sigma$, with $N_c = 1\times10^4$ and $\Delta z = 100$\,$\mu$m. (c) $\mathrm{SD}(\Delta z)$ as a function of the effective HOM interference visibility $V$, controlled by rotating the polarization of one photon using a HWP and with $N_c = 1\times10^4$. The inset shows the measured displacement when $\Delta z$ is set to 50\,$\mu$m for different values of $V$. The blue shaded region indicates the surface flatness tolerance of the HWP ($\lambda/4$ at 633\,nm). (d) Left vertical axis: $\mathrm{SD}(\Delta z)$ as a function of photon loss introduced in one interferometer arm using reflective neutral density filters. Right vertical axis: Standard deviation of the inferred filter thickness after accounting for the refractive index of the filter material (UV Fused Silica, $n=1.4531$ at 810\,nm). The inset compares the filter thickness measured using spectrally resolved HOM interferometry with that measured using a vernier caliper (quoted accuracy of 20\,$\mu$m). The interferometer is initially offset to $\Delta z =-0.302$\,mm, while insertion of the neutral-density filters shifts the measured delay to approximately $+0.27$\,mm, giving a total $\Delta z \approx 0.57$\,mm. The corresponding effective visibilities are $V=0.66$ and 0.70, respectively. The CRB shown corresponds to the differential delay measurement and is obtained by combining the uncertainties associated with the two delay estimates for their respective visibilities. $N_c = 2\times10^4$ when there is zero loss. All error bars of $\mathrm{SD}(\Delta z)$ represent the statistical uncertainty in the measured standard deviation due to the finite number of measurement repeats.}
    \label{Fig5}
\end{figure}

Next, we investigate how the measurement precision depends on several key experimental parameters. Figure~\ref{Fig5}(a) shows the measurement standard deviation $\mathrm{SD}(\Delta z)$ and $\mathrm{SD}(\Delta t)$ as a function of the total number of detected coincidences $N_c$, which includes contributions from both bunched and antibunched photon pairs. The precision improves as $1/\sqrt{N_c}$, consistent with the expected CRB scaling. Figure~\ref{Fig5}(b) shows the dependence of the displacement precision on the RMS spectral bandwidth $\sigma$ of the photon pairs. The effective spectral bandwidth is varied in post-processing by selecting only coincidence events within a specified spectral range. The measurement precision closely follows the prediction obtained from the Fisher information given by Eq.\,\eqref{Fish2}. In both Fig.~\ref{Fig5}(a) and (b), the measurements are taken at $\Delta z = 100$\,$\mu$m, which has a reduced photon effective HOM interference visibility $V=0.91$ due to the reduced ability for the spectrometer resolution to resolve interference fringes of higher oscillation frequencies as discussed earlier.

With the current experimental setup, one second of data acquisition detects approximately $10^4$ photon pairs, resulting in a displacement precision of $\mathrm{SD}(\Delta z)= 45$\,nm, corresponding to a time-delay precision of $\mathrm{SD}(\Delta t)= 150$\,as. With $10^6$ detected photon pairs, the precision improves to $\mathrm{SD}(\Delta z)= 4$\,nm or $\mathrm{SD}(\Delta t)= 13$\,as. Our data analysis software can currently perform the spectral correlation analysis in real time at approximately 1\,Hz for about 2000 coincidence events giving $\mathrm{SD}(\Delta z)= 100$\,nm, $\mathrm{SD}(\Delta t)= 330$\,as. The relatively slow analysis speed is primarily due to the data structure of the TPX3CAM, which requires cluster identification and centroid reconstruction to be performed for every photon detection event~\cite{Zhao2017,Vidyapin2022}. Further details on the data processing are provided in the Supplementary Material, and a demonstration of real-time operation is shown in the Supplementary Video.

We also demonstrate that the interferometer does not require recalibration when the interference visibility of the system changes, as may occur when measuring different samples or transmission channels. In conventional spectrally unresolved HOM interferometry, the interferometer delay must be scanned to characterize the HOM-dip profile. If the interferometer is then operated at a fixed position on the slope of the dip, changes in interference visibility alter the measured coincidence rate and can therefore be indistinguishable from a change in delay. In such cases, the HOM dip must be rescanned to re-establish the coincidence-rate-to-delay response before an accurate delay measurement can be made. Figure~\ref{Fig5}(c) shows the dependence of the measurement precision on $V$, which is controlled by rotating the polarization of one photon using a half wave-plate (HWP). Polarization distinguishability is used here as a convenient and controllable means of varying the visibility; in practice, visibility can also change from other sources of photon distinguishability, such as variations in spatial-mode overlap. As $V$ decreases, the measurement precision degrades in accordance with the theoretical CRB scaling. However, although the reduction in interference visibility degrades the measurement precision, no systematic shift in the estimated displacement is observed over the measured range, with the mean values remaining consistent with the translation-stage positions within the experimental uncertainty, as shown in the inset of Fig.~\ref{Fig5}(c).

Finally, in Fig.~\ref{Fig5}(d) we show the robustness of the method to photon losses, where the thicknesses of different reflective neutral-density filters are measured. As expected, losses increase the measurement uncertainty due to reduced number of detected coincidences. Loss also increases the relative contribution of accidental coincidences in $N_c$, thereby reducing $\eta_{\text{acc}}$ and the effective interference visibility. Following temporal and spectral correlation filtering, the measured coincidence-to-accidental ratios are approximately 57(42) for the antibunched(bunched) channel without added losses, decreasing to 37(26) at the maximum tested loss of approximately 79\%. These correspond to accidental contributions of only approximately 3–4\% of the accepted coincidences. $\eta_{\text{acc}}$ thus remains close to unity and its effect on measurement precision remains relatively small. Consequently, the filter thicknesses remain measurable with good accuracy without requiring interferometer recalibration, demonstrating the robustness of spectrally resolved HOM interferometry against photon loss. We note that conventional HOM interferometry can also compensate for variations in overall photon loss by measuring and normalizing both bunched and antibunched probabilities, as demonstrated in \cite{Colin2025}.

For comparison, before spectral correlation filtering, the corresponding coincidence-to-accidental ratios are only 2.3(1.4) without additional loss and 1.0(0.4) at the maximum tested loss. The substantial improvement following spectral filtering highlights the effectiveness of the strong frequency anticorrelation of the photon pairs in rejecting uncorrelated background events~\cite{Zhang2020}.

The measurements shown in Figs.\,\ref{Fig3}, \ref{Fig4} and \ref{Fig5}(a, b) were acquired on the same day under the same experimental conditions. The measurements in \ref{Fig5}(c, d) were performed approximately two months later using the same experimental setup and nominally the same operating conditions. These results indicates the reproducibility of the measurement over an extended period. However, the experiment was not designed as a systematic study of long-term stability or environmental sensitivity, and parameters such as temperature and mechanical drift were not independently varied or controlled for this purpose.

We note that the two-photon probability distribution given by Eq.\eqref{Cprob} is even in the delay, such that the fringe periodicity determines only the magnitude $|\Delta t|$ and cannot distinguish its sign. In the measurements presented here, the sign of $\Delta t$ is known from the direction of motion of the translation stage and is assigned accordingly. This is sufficient for applications such as thickness measurements, where the direction of the optical-path change is known a priori. For a general displacement sensor requiring independent determination of both magnitude and direction, an additional reference would be required; for example, the interferometer could be operated about a known nonzero delay so that small positive and negative displacements are inferred relative to this reference, provided the displacement does not cross the zero-delay point.

In the present experiment the interferometer is not actively stabilized, and slow environmental fluctuations such as mechanical vibrations and thermal drift can introduce low-frequency variations in the interferometer path length during long acquisitions. To suppress the influence of these fluctuations, we record a long continuous dataset of photon-detection events and randomly shuffle the detected photon pairs before dividing them into non-overlapping measurement batches. Each coincidence event is assigned to only one batch and is not reused in any other repetition. The shuffling therefore randomizes the temporal ordering of the events and averages over slow interferometer fluctuations. The resulting distribution of delay estimates across the independent batches is used to determine the statistical uncertainty of a single measurement containing the same number of detected photon pairs. As shown in the Supplementary Material, the unshuffled data begin to deviate from the shuffled data for $N_C\gtrsim2000$, corresponding to acquisition times of $\sim$0.2\,s, indicating the onset of drift-induced degradation. Importantly, the real-time demonstration uses a 0.2\,s acquisition time with $N_C\sim2000$ photons per acquisition and therefore do not require random shuffling. Each 0.2\,s acquisition is processed in approximately 1\,s, with acquisition of the subsequent batch performed concurrently with processing, yielding an effective real-time update rate of $\sim1$\,Hz.

Lastly, we note that the uncertainties reported throughout this section, including the error bars in Figs.\,\ref{Fig3}-\ref{Fig5}, quantify only the statistical precision of the spectrally resolved HOM delay measurement. The systematic uncertainty associated with the spectrometer wavelength calibration is not included. Once the spectrometer is calibrated, this calibration error remains fixed for subsequent measurements performed using the same wavelength mapping and therefore does not contribute to the shot-to-shot variation in the inferred delay. Based on an estimated wavelength-calibration uncertainty of approximately 0.02\,nm over the 79.98\,nm measured spectral range, this corresponds to a systematic scale uncertainty of up to approximately 0.025\% in the measured displacement. For practical metrology applications, this calibration uncertainty should be included in the overall uncertainty budget. See the Supplementary on details of the spectrometer calibration.

\section{Discussion}

We have demonstrated real-time spectrally resolved Hong–Ou–Mandel (HOM) interferometry using broadband SPDC photons as a practical platform for quantum-enhanced displacement sensing over an extended dynamic range. By resolving the joint spectral correlations of entangled photon pairs and extracting the time delay from the periodicity of the spectral interference fringes, the method avoids the conventional limitation imposed by the photon coherence length (HOM dip width) that constrains traditional HOM dip measurements. As a result, we show near-optimal precision can be maintained for path differences two orders of magnitude larger than the HOM dip width. Using $10^6$ detected photon pairs, we experimentally achieve a displacement precision of approximately 4\,nm (13\,as), and real-time operation with our current detection equipment yields a precision of about 100\,nm (330\,as). In addition, because the measurement relies on fringe periodicity rather than absolute coincidence rates, it is intrinsically robust to variations in interference visibility, eliminating the need for recalibration which is typically required in conventional HOM metrology. Further improvements in precision can be achieved by using broadband non-degenerate SPDC photon pairs, which introduces a non-zero $\Omega$ in Eq.\,\eqref{Fish2}. 

More broadly, the ability to perform high-precision delay estimation without scanning the interferometer or recalibrating the HOM dip makes spectrally resolved HOM interferometry particularly attractive for real-world sensing applications. Potential directions include precision thickness metrology of transmissive materials, quantum optical coherence tomography, and long-range displacement sensing in environments where losses and visibility fluctuations are unavoidable. With continued advances in photon-pair sources and high-efficiency detector arrays, spectrally resolved HOM interferometry offers a promising route toward practical quantum-enhanced metrology operating at the ultimate precision limits set by quantum mechanics.

It is useful to compare the present implementation with the recent frequency-resolved HOM delay measurement of \cite{Brooks2026}. Both approaches extend the measurement range beyond the conventional HOM interference window by extracting delay from spectrally resolved two-photon interference, but operate in different regimes. Reference \cite{Brooks2026} used independent, spectrally uncorrelated photons at telecom wavelengths together with a time-of-flight spectrometer based on approximately 50\,km of dispersive fiber. In contrast, the present experiment uses broadband frequency-anticorrelated SPDC photon pairs and a free-space grating spectrometer. For the same marginal spectral bandwidth, perfect frequency anticorrelation provides a factor-of-two increase in the ideal Fisher information relative to spectrally uncorrelated photons, while the broader spectral bandwidth further improves the achievable precision. Moreover, the strong spectral anticorrelation enables substantially improved rejection of uncorrelated background through joint spectral filtering, providing an additional practical advantage in noisy measurement environments. Reference \cite{Brooks2026} reports a displacement precision of approximately 150\,nm (0.5\,fs) using $7.9\times10^6$ detected events with a maximum measurement range of approximately $\pm18$\,mm, approximately 20 times its photon coherence length. In comparison, this experiment achieves a displacement precision of 4\,nm (13\,as) with approximately $10^6$ detected photon pairs with a maximum range of $\pm1$\,mm, over 350 times the photon coherence length. Thus, while \cite{Brooks2026} offers a larger absolute measurement range, our approach achieves substantially higher precision and a larger range relative to the photon coherence length, together with improved background rejection and real-time operation at approximately 1\,Hz.

The time-tagging-camera-based two-photon spectrometer implemented with the TPX3CAM has already enabled a range of proof-of-principle quantum-optical measurement applications, including quantum snapshot hyperspectral imaging~\cite{Zhang2023}, spectrally resolved quantum illumination~\cite{Zhang2020}, and spectroscopy based on time and wavelength resolved photon correlations~\cite{Farella2024,Jordan2026}. Despite this versatility, the performance of the present system, particularly its real-time measurement precision and update rate, is strongly limited by the photon-detection efficiency of the TPX3CAM, which is only $\sim$8\% at the operating wavelength~\cite{Vidyapin2022}. Consequently, the present real-time operation should be regarded as a proof-of-principle demonstration of the capability of the platform, with substantially higher-efficiency detectors required to approach the performance needed for practical sensing applications. Emerging detector technologies, including SPAD and superconducting nanowire single-photon detector (SNSPD) arrays~\cite{Wollman2019,Steinhauer2021,Oripov2023}, offer substantially higher detection efficiencies and lower timing jitter than the present TPX3CAM. Recent frequency-resolved HOM measurements using an 8-element SPAD array~\cite{DiLena2026} as well as integrated photon-counting spectrometer based on a 16-element SNSPD array~\cite{Zheng2023} has been demonstrated. Although these implementations currently provide substantially fewer independent spectral channels than the $\sim$307 achieved by the present TPX3CAM spectrometer, continuing development of large-format detector arrays, including SNSPD cameras containing hundreds of thousands of pixels~\cite{Oripov2023}, provides a promising route toward substantially finer spectral resolution and/or broader simultaneous spectral coverage. In addition, SNSPD arrays can achieve quantum efficiencies exceeding 80\%, which would increase the detected coincidence rate by more than two orders of magnitude compared with the present system. Under otherwise identical conditions, this improvement would enable the acquisition of $10^6$ detected photon pairs in approximately one second. Their substantially lower timing jitter would also permit a narrower coincidence window, thereby reducing accidental and background coincidences and improving the background-limited signal-to-noise ratio. Thus, such detector developments could substantially improve the spectral resolution, measurement precision, background rejection, and overall practical applicability of the demonstrated approach.

\bibliography{SHOMref}

\section*{Supplementary}

\section*{Detailed experimental setup}
\subsection*{Experimental Setup} 
\begin{figure}
    \centering
    \includegraphics[width=1\linewidth]{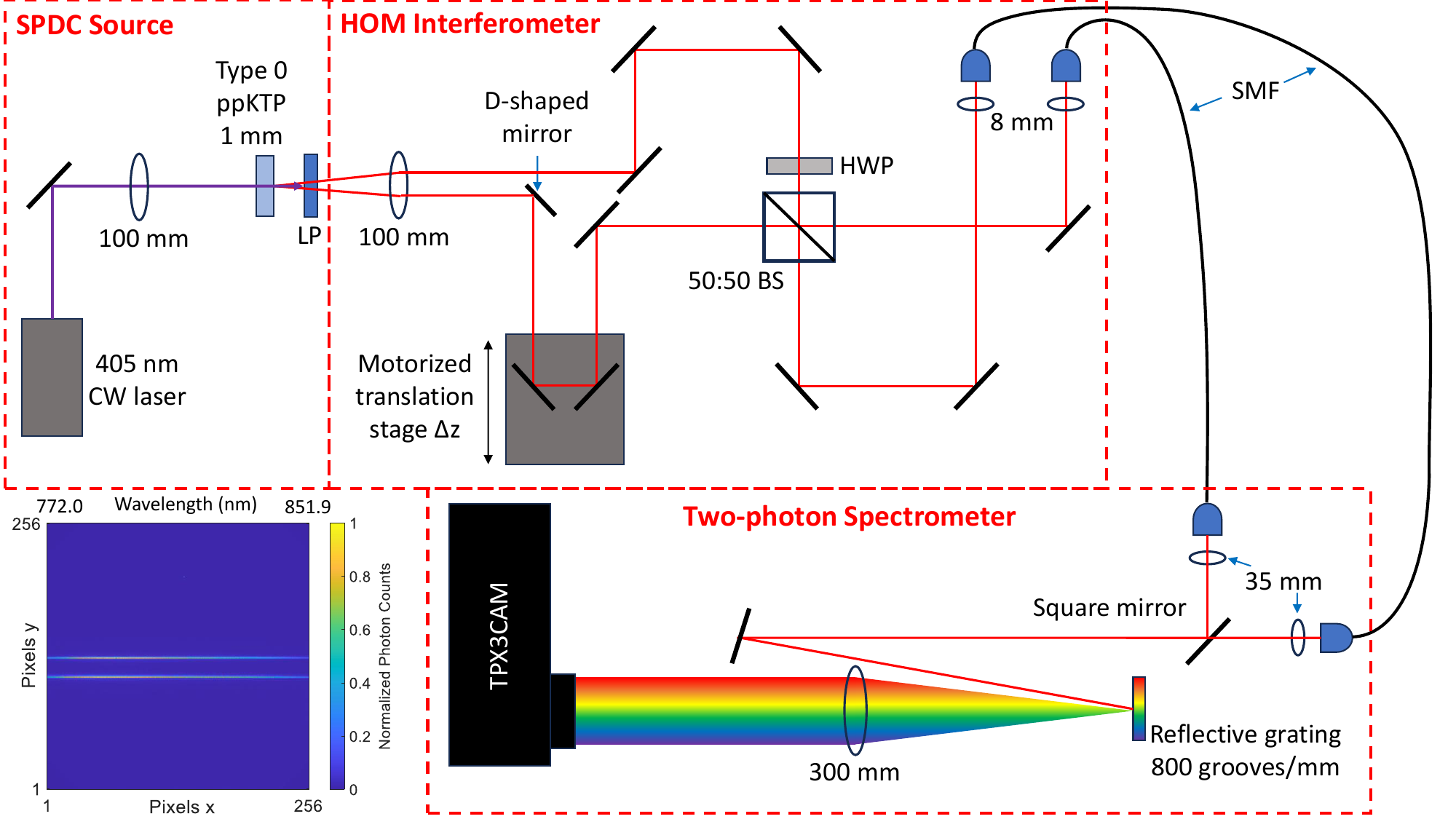}
    \caption{{\bf Experimental setup for spectrally resolved HOM interferometry.} LP: long-pass filter, HWP: half-wave plate, BS: beamsplitter, SMF: single-mode fiber. Bottom left corner shows the raw twin-beam spectrum recorded by the camera.}
    \label{Setup}
\end{figure}

Figure~\ref{Setup} shows the detailed experimental setup for spectrally resolved HOM interferometry. A 40\,mW 405.05\,nm CW laser (Cobolt 08-NLD) with a 1\,pm linewidth pumps a 1\,mm-thick Type-0 periodically poled potassium titanyl phosphate (ppKTP) crystal to generate broadband spectrally entangled photon pairs via SPDC. The pump beam is focused onto the crystal using a 100\,mm lens to reduce the number of generated SPDC spatial modes and improve fiber coupling efficiency.

The generated photon pairs are spatially separated into two paths using a D-shaped mirror and subsequently recombines at a 50:50 BS to undergo HOM interference. A motorized translation stage in one arm of the interferometer provides precise control of the relative optical delay between the photons. A HWP placed in the other path adjusts the photon polarization to control the interference visibility.

After the beamsplitter, photons from the two output ports are coupled into SMFs and directed to a custom-built two-photon spectrometer based on the TPX3CAM~\cite{Nomerotski2019,ASI2}). Within the spectrometer, the photons are dispersed by an 800 grooves/mm reflective grating, and the first diffraction order is imaged onto the camera sensor. The two SMF outputs are vertically offset so that photons from each output port of the interferometer are mapped onto separate regions of the camera. 

Within the spectrometer, the photons are dispersed by an 800 grooves/mm reflective grating. Due to the large camera pixel size of 55\,$\mu$m and $256\times256$ pixel array, a 300\,mm lens is used to focus the spectrum onto the camera so that all pixels are filled.

\subsection*{Spectrometer Resolution}
The SMF used at the spectrometer input has a core diameter of $5.3$\,$\mu$m. The fiber output is magnified by a factor of 8.6 onto the camera sensor, which has a width of 14\,mm. This configuration corresponds to $\frac{14}{8.6\times 0.0053}\approx307$ independent spectral elements across the detector. To adequately sample this resolution according to the Nyquist criterion, approximately three pixels per spectral element are required, corresponding to about 921 pixels across the sensor.

The event camera has a native resolution of $256\times256$ pixels. By applying a centroiding algorithm to the detected photon clusters generated through the image intensifier~\cite{Kim2020}, the effective spatial resolution is increased to $1024\times1024$ pixels (see Section II for details). This enables the spectrometer to operate at its theoretical resolution limit.

Experimentally, the measured spectrum spans from 772.08\,nm to 852.06\,nm across the camera sensor. With 307 spectral elements, this corresponds to a spectral resolution of $\frac{852.06-772.08}{307} \approx 0.26$\,nm. The 0.26\,nm resolution should be regarded as a property of the complete spectrometer rather than a fundamental limit of the TPX3CAM. For broadband spectroscopy, a more useful figure of merit is the number of independently resolvable spectral channels, since finer absolute spectral resolution can generally be achieved at the expense of a reduced simultaneously accessible spectral range, for example by increasing the grating dispersion or imaging focal length. For a detector of fixed active width, increasing the dispersion therefore primarily trades spectral range for spectral resolution rather than increasing the number of independent spectral channels. Increasing the detector dimension along the dispersion direction, on the other hand, would allow more independent spectral channels, enabling finer resolution without sacrificing spectral range, or a broader spectral range at the same resolution. Further improvement in the number of spectral channels would therefore require a sensor with more pixels along the dispersion direction.

From Eq.(2) of the main text we see that for a phase change of $2\pi$, corresponds to a frequency change of $\Delta\omega=\pi/\Delta t$. Converting to displacement and wavelength gives 
\begin{equation}
    \Delta z = \frac{\lambda^2}{2\Delta \lambda}
\end{equation} 
So, with $\Delta \lambda = 0.26$\,nm and using the shortest wavelength in the detected band of $\lambda=772$\,nm gives an upper limit of $|\Delta z| \approx 1.15$\,mm. On the other hand, for the lower limit $\Delta \lambda = 80$\,nm, which would give a $|\Delta z| \approx 3.7$\,$\mu$m.

\section*{TPX3CAM data processing}
\subsection*{Raw data processing}
The TPX3CAM is an event-based camera system in which each detected photon is assigned a time stamp and pixel address. Its raw data output therefore consists of a list of numbers containing the time and pixel for each detected event rather than conventional image frames. The sensor is silicon-based with a $256\times256$ pixel array, and each pixel provides timing information with a resolution of 1.6\,ns. Achieving per-pixel timing requires a comparatively large pixel size 55\,$\mu$m to accommodate the in-pixel circuitry, which in turn leads to a relatively high noise floor that prevents direct single-photon sensitivity.

Single-photon detection is enabled by coupling the camera to an image intensifier. A photon striking the photocathode ejects an electron, which is amplified through a microchannel plate into a cascade of electrons that excite a phosphor screen. The resulting light flash is sufficiently bright to be detected by the camera, typically illuminating a small cluster of pixels. These clusters are reconstructed into single-photon events using a centroiding algorithm.

For each pixel in a cluster, the camera records a time-of-arrival (TOA), defined as the moment when the signal first crosses the discrimination threshold. The camera also measures a time-over-threshold (TOT), the duration for which the signal remains above this threshold. Because a larger photon flux causes the signal to cross the threshold earlier and to remain above it longer, the TOA and TOT are thus interdependent: events with larger TOT values exhibit earlier TOAs. As the local photon flux varies across the pixels within a cluster, both TOT and TOA vary accordingly, even though all pixels originate from the same photon and should share a common true arrival time. To correct this, a TOA–TOT correlation is extracted from large datasets and applied in two steps: first, to unify the TOA across all pixels within each individual cluster; and second, to correct residual TOA offsets between different clusters arising from the fact that the TOT range (and thus the TOA bias) differ from cluster to cluster. After applying this timing correction, a timing resolution of approximately 8\,ns is achieved, still above the 1.6\,ns pixel resolution, but an order of magnitude better than the uncorrected cluster timing. Further details of the raw data processing are provided in Refs.\,\cite{Nomerotski2019,Vidyapin2022}.

The final processed data are stored as an $N\times3$ matrix, where $N$ is the number of detected photon events, and the three columns correspond to the corrected TOA and the $x$ and $y$ pixel coordinates of each event.

\subsection*{Pixel super-resolution through centroiding}
\begin{figure}
    \centering
    \includegraphics[width=1\linewidth]{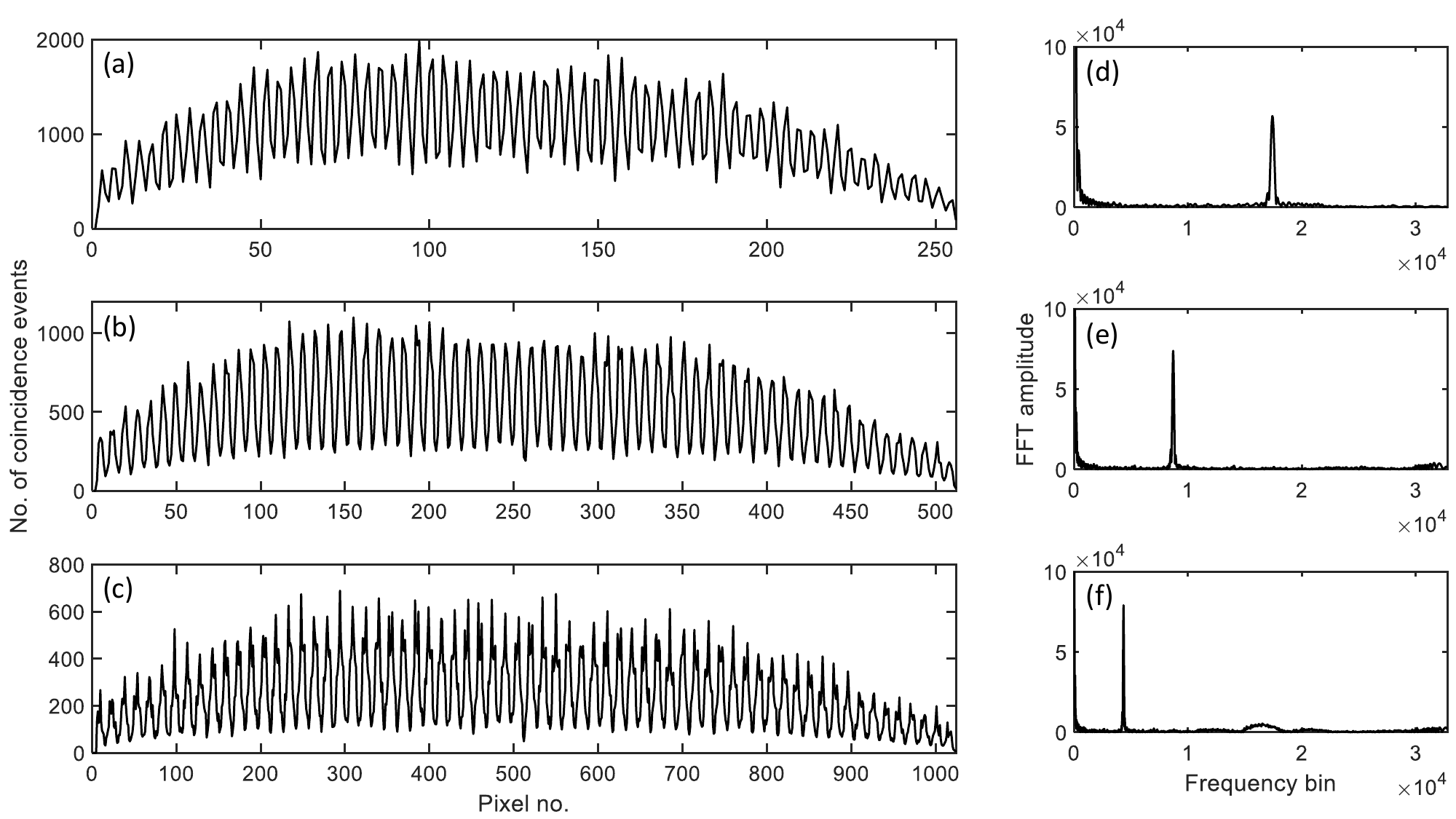}
    \caption{Cross-section of the two-photon interference spectral pattern at $\Delta z = 280$\,$\mu$m with $3\times10^5$ coincidence events shown for (a) $256$ pixels - native camera resolution. (b) $512$ pixels  and (c) $1024$ pixels. (d-f) shows the corresponding FFT spectrum (only positive frequency component shown) performed at a resolution of 65536 points through zero padding of the measured data.}
    \label{Supp1}
\end{figure}

Since each photon event is detected as a cluster of pixels with the local pixel photon flux given by the TOT, the photon position can be localized with sub-pixel precision using a center-of-mass estimator,
\begin{equation}
\langle x \rangle = \frac{\sum_i x_i \text{TOT}_i}{\sum_i \text{TOT}_i}.
\end{equation}
This enables effective pixel super-resolution beyond the native camera grid of $256\times256$ pixels~\cite{Kim2020}. The impact of this approach is illustrated in Fig.\,\ref{Supp1}, which shows the cross-section of the two-photon spectral interference pattern reconstructed at different effective pixel resolutions of 256 (native), 512, and 1024 pixels. As the resolution increases, the interference fringes visibility improves allowing for more accurate estimation of high-frequency oscillations. More importantly, the enhanced sampling provided by sub-pixel localization allows access to higher spatial (spectral) frequency oscillations that would otherwise be undersampled or aliased at the native camera resolution.

\subsection*{Coincidence processing}
For coincidence detection between two regions of the camera, either between the top and bottom spectral band or between the short and long wavelength of the same spectral band, the full $N\times3$ processed data matrix is first separated into two submatrices, $N_s\times3$ and $N_i\times3$ matrices (with $N_s+N_i=N$). Each submatrix contains only the events whose $(x,y)$ pixel coordinates fall within the corresponding spatial region of interest.

Coincidence events are then identified by comparing the TOA values in the two submatrices. A pair of events is classified as a coincidence if their TOAs fall within a predefined temporal window centered on the correlation peak of the biphoton arrival-time histogram (20\,ns in this case). Events that do not satisfy this temporal criterion are discarded as accidentals or background.

Once coincidence pairs are identified, the TOA information is no longer required. The resulting coincidence dataset is stored as an $N_c\times4$ matrix, where $N_c$ is the number of coincidence events, and the four columns correspond to the $(x_s,y_s)$ pixel coordinates of the signal photon and the $(x_i,y_i)$ pixel coordinates of the idler photon detected in coincidence. This coincidence list serves as the input for constructing the two photon spectral joint distribution.

\subsection*{Spectrometer Calibration}
The spectrometer wavelength calibration was performed in two stages. First, the pump wavelength of 405.05\,nm, known from the manufacturer's calibration, was used to constrain the central signal and idler frequencies through energy conservation. An initial pixel-to-wavelength mapping was established by matching the measured spectral-fringe periodicity to the theoretically expected values at known interferometer delays. For finer calibration, the interferometer delay was then scanned from $-460$\,$\mu$m to $+460$\,$\mu$m in 20\,$\mu$m steps (ignoring zero displacement due to the fringe being unresolvable). The wavelength mapping of the idler spectral axis was adjusted until the mean displacement step inferred from the spectral fringes matched the applied 20\,$\mu$m step. Averaging over the full $920$\,$\mu$m calibration range reduces the influence of the finite repeatability of the translation stage.

From the range of wavelength mappings that are consistent with the measured displacement steps, we estimate a wavelength-calibration uncertainty of approximately 0.02\,nm over the 79.98\,nm spectral range. This corresponds to a conservative fractional scale uncertainty of approximately 0.025\%, equivalent to systematic displacement uncertainties of approximately 25\,nm at $\Delta z = 100$\,$\mu$m and 100\,nm at $\Delta z = 400$\,$\mu$m. Since this calibration uncertainty remains fixed for subsequent measurements using the same wavelength mapping, it does not contribute to the statistical spread of the measured displacements and is therefore not included in the reported error bars.

We note that this procedure calibrates the overall wavelength mapping rather than independently determining the wavelength associated with each detector pixel. Residual nonlinearities or pixel-to-pixel variations in the spectrometer dispersion may locally distort the fringe periodicity and contribute to the measurement uncertainty. For applications requiring traceable absolute displacement accuracy, a more rigorous pixel-by-pixel calibration using known spectral lines would be necessary.

\section*{Comparison and combined analysis of bunched and anti-bunched detection channels}

The bunched and anti-bunched detection channels exhibit complementary spectral interference fringes, which nevertheless produce FFT peaks at the same absolute delay. To assess the influence of the dark band in the bunching spectrum, arising from the inability of the camera to resolve two closely spaced photon events, we compared the delay estimates obtained from the two channels separately.

\begin{figure}
    \centering
    \includegraphics[width=1\linewidth]{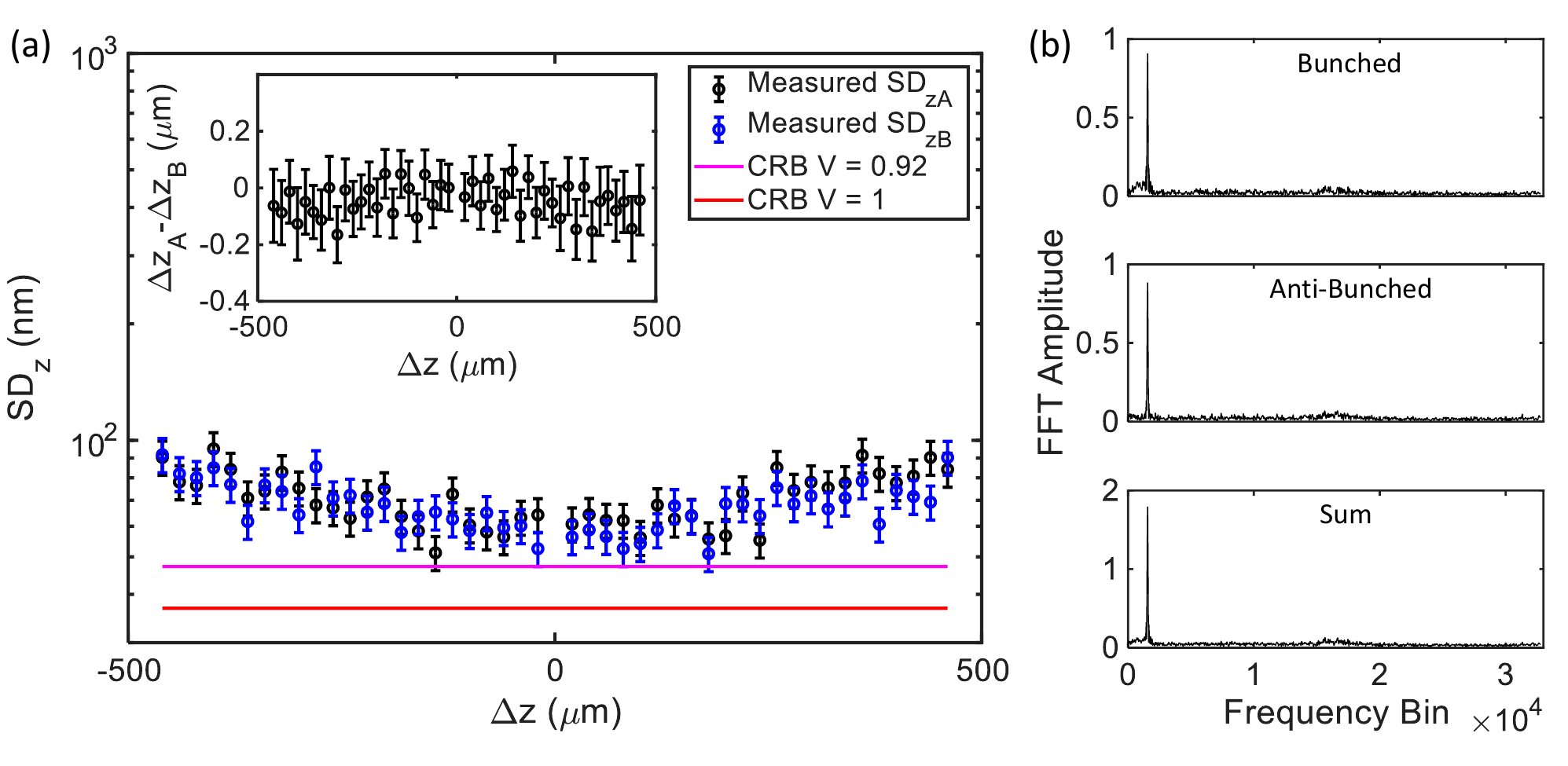}
    \caption{Comparison of the bunched and anti-bunched detection channels. (a) Standard deviation in the measured $\Delta z$, calculated separately using bunched (blue) and anti-bunched (black) coincidence events with a total $N_c=10^4$. Inset shows the difference in the measured $\Delta z$. (b) Corresponding FFT magnitude profiles at $\Delta z = 100$\,$\mu$m for bunched, anti-bunched and their sum. Note that the spectral envelope was subtracted from each spectrum before performing the FFT to remove the zero-frequency peak.}
    \label{Supp2}
\end{figure}

For datasets containing $10^4$ detected coincidences, the average numbers of bunched and anti-bunched events were 4990.5 and 5009.5, respectively. As shown in Fig.\,\ref{Supp2}(a), the two channels exhibit comparable delay-estimation uncertainties across the measured displacement range, indicating that the missing spectral region in the bunching channel has no significant effect on the measurement precision. The  Furthermore, the FFT profiles between the two channels at $\Delta z = 100$\,$\mu$m are compared in Fig.\,\ref{Supp2}(b), indicating no discernible channel-dependent shift in the estimated delay. Averaged over all measured values of $\Delta z$, the mean difference in the inferred displacement between the anti-bunched and bunched channels was $-0.045 \pm 0.059$\,$\mu$m, consistent with no significant channel-dependent displacement bias.

In our analysis, the FFTs of the bunched and anti-bunched spectra are calculated separately, and their resulting magnitude profiles are added before determining the dominant peak position. 

\section*{Gaussian spectrometer response and delay-dependent Fisher information}
\subsection*{Visibility reduction factor due to spectrometer response}
The finite spectral resolution of the spectrometer reduces the visibility of the measured spectral interference fringes. We model this effect by convolving the ideal spectral interference pattern with a Gaussian spectrometer point-spread function (PSF).

For perfectly energy-anticorrelated photon pairs, the photon frequencies satisfy $\omega_1+\omega_2=\omega_p$. The interference term in Eq.(2) of the main text can therefore be written as
\begin{equation}
    \cos[(\omega_1-\omega_2)\Delta t] = \cos[(2\omega_1-\omega_p)\Delta t].
\end{equation}
We approximate the spectrometer response by a normalized Gaussian PSF in the spectral coordinate $\omega_1$
\begin{equation}
    R(\delta\omega)= \frac{1}{\sqrt{2\pi}\sigma_{\mathrm{PSF}}} \exp\left[-\frac{\delta\omega^2}{2\sigma_{\mathrm{PSF}}^2}\right],
\end{equation}
where $\sigma_{\mathrm{PSF}}$ is the RMS angular-frequency width of the spectrometer response.

\begin{figure}
    \centering
    \includegraphics[width=0.6\linewidth]{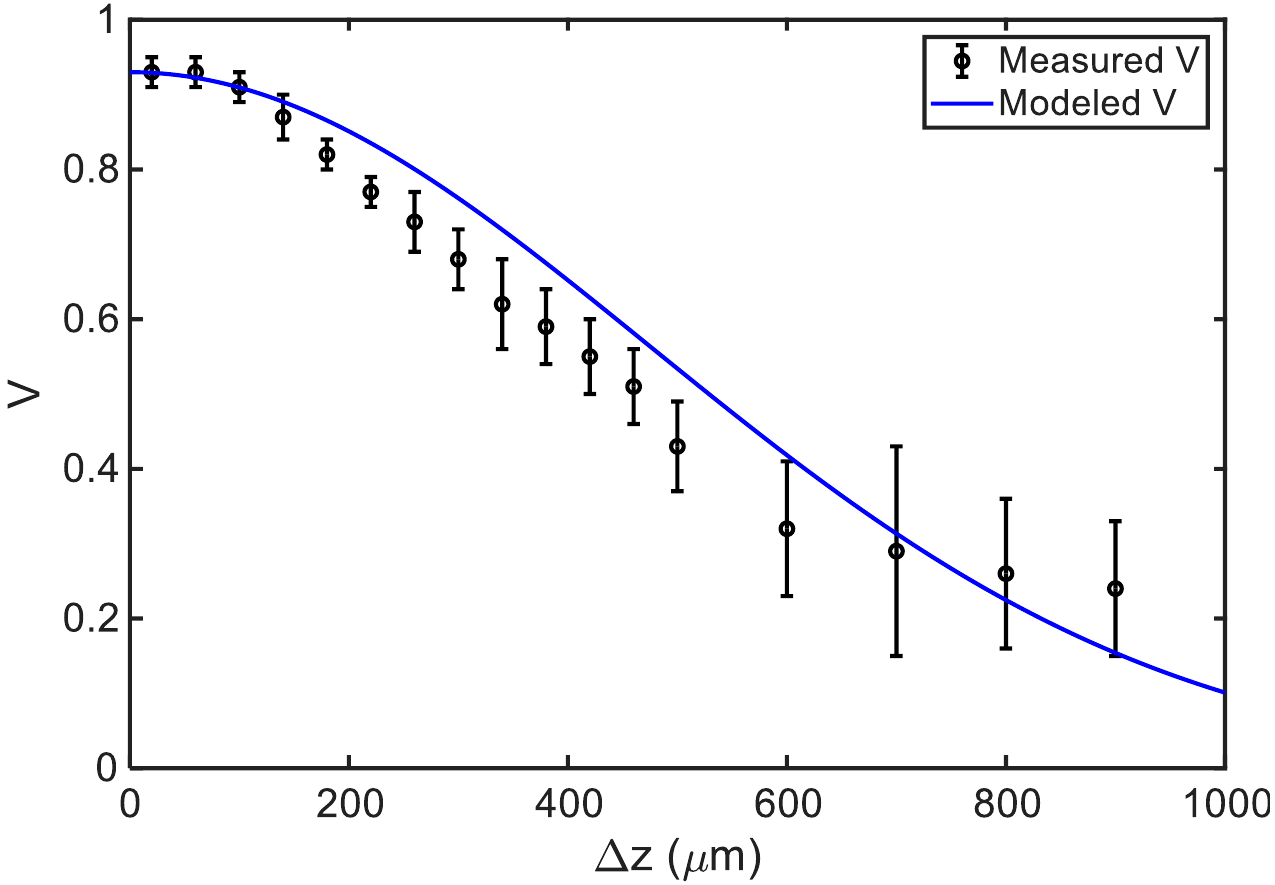}
    \caption{Comparison between the experimentally measured $V(\Delta z)$ against the modeled $V(\Delta z) = \eta^2\xi(\Delta z)$ with $\eta^2 = 0.92$ and $\xi(\Delta z)$ given by Eq.\eqref{SPSF2}.}
    \label{Supp3}
\end{figure}

The convolution of the interference term with the Gaussian PSF gives
\begin{align}
    I(\omega_1,\Delta t) &= \int_{-\infty}^{\infty} R(\delta\omega)\cos[(2(\omega_1-\delta\omega)-\omega_p)\Delta t]\,d\delta\omega\nonumber\\
    &= \cos[(2\omega_1-\omega_p)\Delta t]\int_{-\infty}^{\infty}R(\delta\omega)\cos(2\delta\omega\Delta t)\,d\delta\omega\nonumber\\
    &\quad + \sin[(2\omega_1-\omega_p)\Delta t]\int_{-\infty}^{\infty}R(\delta\omega)\sin(2\delta\omega\Delta t)\,d\delta\omega\nonumber\\
    &= \cos[(2\omega_1-\omega_p)\Delta t] \exp(-2\sigma_{\mathrm{PSF}}^2\Delta t^2).
\end{align}
Consequently, this gives the fringe-visibility reduction factor
\begin{equation}
    \xi(\Delta t)=\exp(-2\sigma_{\mathrm{PSF}}^2\Delta t^2),
    \label{SPSF}
\end{equation}
or converting to wavelength and displacement $\Delta z$ gives
\begin{equation}
    \xi(\Delta z)=\exp\left(-\frac{8\pi^2\sigma_{\lambda}^2\Delta z^2}{\lambda_0^4}\right).
    \label{SPSF2}
\end{equation}
For this experiment $\lambda_0 = 810$\,nm and $\sigma_{\lambda} = 0.11$\,nm for a FWHM spectral resolution of 0.26\,nm. A plot comparing the experimentally measured $V(\Delta z)$ at various delays with that determined using $V(\Delta z) = \eta^2\xi(\Delta z)$ is shown in Fig,~\ref{Supp3}.

The measured interference fringe visibility is determined using
\begin{equation}
V(\Delta z) = \frac{1}{n}\sum_{i=1}^{n}\frac{N_{i\max}-N_{i\min}}{N_{i\max}+N_{i\min}},
\end{equation}
where $N_{i\max}$ and $N_{i\min}$ are the $i$th maxima and minima of the measured spectral fringes identified automatically using the LabVIEW peak-detection routine, which fits a quadratic polynomial to sequential groups of data points to determine the local extrema.

\subsection*{Delay-dependent Fisher information}
Expanding Eq.(5) of the main text we get ($\Delta t$ dependence of $V$ is omitted for notational simplicity)
\begin{align}
\mathcal{F}(\Delta t) &= V^2\iint |\Phi|^2\frac{\Delta\omega^2\sin^2\phi}{1-V^2\cos^2\phi}\,d\omega_1d\omega_2 \nonumber\\
&\quad + (V')^2\iint |\Phi|^2\frac{\cos^2\phi}{1-V^2\cos^2\phi}\,d\omega_1d\omega_2 \nonumber\\
&\quad - 2VV'\iint |\Phi|^2\frac{\Delta\omega\sin\phi\cos\phi}{1-V^2\cos^2\phi}\,d\omega_1d\omega_2,
\end{align}
where $\phi = \Delta\omega\Delta t$. For notational simplicity, the explicit $\Delta t$ dependence of the effective visibility $V$ is omitted, with $V$ understood to represent $V(\Delta t)$ and $V' = dV/d\Delta t$.

In the large-delay regime, the spectral fringes oscillate rapidly across the bandwidth and the averaging the oscillatory terms gives
\begin{align}
& \left\langle \frac{V^2\sin^2\phi)}{1-V^2\cos^2\phi} \right\rangle_\phi = 1-\sqrt{1-V^2} \nonumber\\
&\left\langle \frac{\cos^2\phi}{1-V^2\cos^2\phi} \right\rangle_\phi = \frac{1}{V^2}\left(\frac{1-\sqrt{1-V^2}}{\sqrt{1-V^2}}\right) \nonumber\\
&\left\langle \frac{\sin\phi\cos\phi}{1-V^2\cos^2\phi} \right\rangle_\phi = 0.
\end{align}
Consequently, 
\begin{align}
\mathcal{F}(\Delta t) &\approx\left(1-\sqrt{1-V^2}\right)\int\int d\omega_1d\omega_2 \left|\Phi(\omega_1,\omega_2)\right|^2\Delta\omega^2 \nonumber\\
&\quad + \frac{(V')^2}{V^2}\left[\frac{1-\sqrt{1-V^2}}{\sqrt{1-V^2}}\right] \int\int d\omega_1d\omega_2 \left|\Phi(\omega_1,\omega_2)\right|^2\nonumber\\
&= \left(1-\sqrt{1-V^2}\right)\left[ \left\langle \Delta\omega^2 \right\rangle + \frac{(V')^2}{V^2\sqrt{1-V^2}}\right] .
\end{align}
For perfectly frequency-anticorrelated photons with equal RMS spectral bandwidth $\sigma$ 
\begin{equation}
    \left\langle \Delta\omega^2 \right\rangle = 4\sigma^2 + \Omega^2,
\end{equation}
and using the Gaussian spectral PSF from Eq.\eqref{SPSF} gives
\begin{equation}
    \mathcal{F}(\Delta t) \approx \left(1-\sqrt{1-V^2}\right)\left[4\sigma^2 + \Omega^2 + \frac{16\sigma_{\mathrm{PSF}}^4\Delta t^2}{\sqrt{1-V^2}}\right].
\end{equation}
Since for our experiment $\sigma \gg \sigma_{\mathrm{PSF}}$, the contribution from the last term is strongly suppressed therefore giving us
\begin{equation}
    \mathcal{F}(\Delta t) \approx \left(1-\sqrt{1-V^2}\right)(4\sigma^2 + \Omega^2).
\end{equation}

\section*{BBO crystal thickness measurement}
\subsection{Path length difference in propagation through BBO crystal}

\begin{figure}
    \centering
    \includegraphics[width=0.5\linewidth]{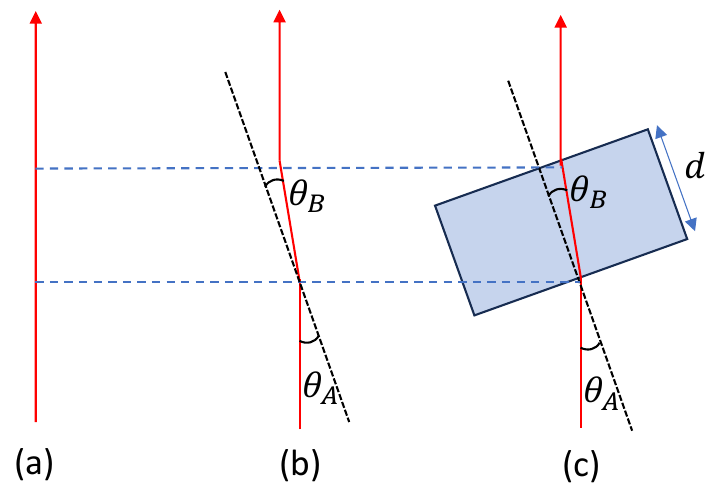}
    \caption{Illustration of the change in path length as light enters a crystal of thickness $d$. }
    \label{Supp4}
\end{figure}

The extra path from the beam deviation between Fig.\,\ref{Supp4}(a) and (b) is
\begin{align}
    \Delta z_1 &= \frac{n_Ad}{\cos\theta_B} - \frac{n_Ad}{\cos\theta_B}\cos\left(\theta_A-\theta_B\right)\nonumber\\
             &= \frac{n_Ad}{\cos(\theta_An_A/n_B)}\left[1 - \cos\left(\frac{\theta_A(n_A-n_B)}{n_B}\right)\right],
\end{align}
where we have used Snell's law $n_A\sin\theta_A = n_B\sin\theta_B$ and the small angles approximation $\lim_{\theta\rightarrow0}\sin\theta\approx\theta$. 

The extra path due to the increased refractive index between Fig.\,\ref{Supp4}(b) and (c) is
\begin{align}
    \Delta z_2 &= \frac{n_Bd}{\cos\theta_B} - \frac{n_Ad}{\cos\theta_B} \nonumber\\
               &= \frac{d(n_B-n_A)}{\cos(\theta_An_A/n_B)}.
\end{align}

Thus the total increase in path length is
\begin{align}
    \Delta z &= \Delta z_1 + \Delta z_2 \nonumber\\
             &= \frac{d}{\cos(\theta_An_A/n_B)}\left[n_B - n_A\cos\left(\frac{\theta_A(n_A-n_B)}{n_B}\right)\right].
\end{align}

\subsection{Refractive-index correction and uncertainty}

The physical thickness of the BBO crystal was determined from the group delay introduced by its insertion into the optical path. Because the measurement is sensitive to the group delay, the relevant refractive index is the group refractive index, $n_g = n_o(\lambda) - \lambda(dn_o/d\lambda)$, rather than the ordinary phase refractive index $n_o$.

For the primary analysis, we used the ordinary-ray Sellmeier relation reported by Zhang et al.~\cite{Zhang2000}
\begin{equation}
    n_o^2(\lambda) = 2.7359 + \frac{0.01878}{\lambda^2 - 0.01822} - 0.01471\lambda^2 + 0.0006081\lambda^4 - 0.00006740\lambda^6,
    \label{Sell}
\end{equation}
where $\lambda$ is expressed in $\mu$m. Evaluating the corresponding group refractive index at the degenerate photon wavelength of 810\,nm gives $n_g = 1.6839$. Using this value, the thickness inferred from the measured delay is $d_0 = 312.85 \pm 0.03$\,$\mu$m.

The uncertainty associated with the refractive-index model was examined by comparing this result with the Sellmeier relation of Eimerl et al.~\cite{Eimerl1987}
\begin{equation}
    n_o^2(\lambda) = 2.7405 + \frac{0.0184}{\lambda^2 - 0.0179} - 0.0155\lambda^2,
\end{equation}
which gives $n_g = 1.6850$ at 810\,nm. Applying this alternative index to the same measured group delay gives $d_0 = 312.35\pm 0.03$\,$\mu$m.

\subsection{Independent classical spectrally resolved interferometry}

To independently verify the thickness measurement, classical spectrally resolved first-order interferometry was performed using a Coherent Chameleon Ultra Titanium-sapphire femtosecond laser. The laser power was attenuated to approximately 100\,$\mu$W, and the interference spectra were recorded using an Ocean Optics HR2000+ spectrometer.

\begin{figure}
    \centering
    \includegraphics[width=1\linewidth]{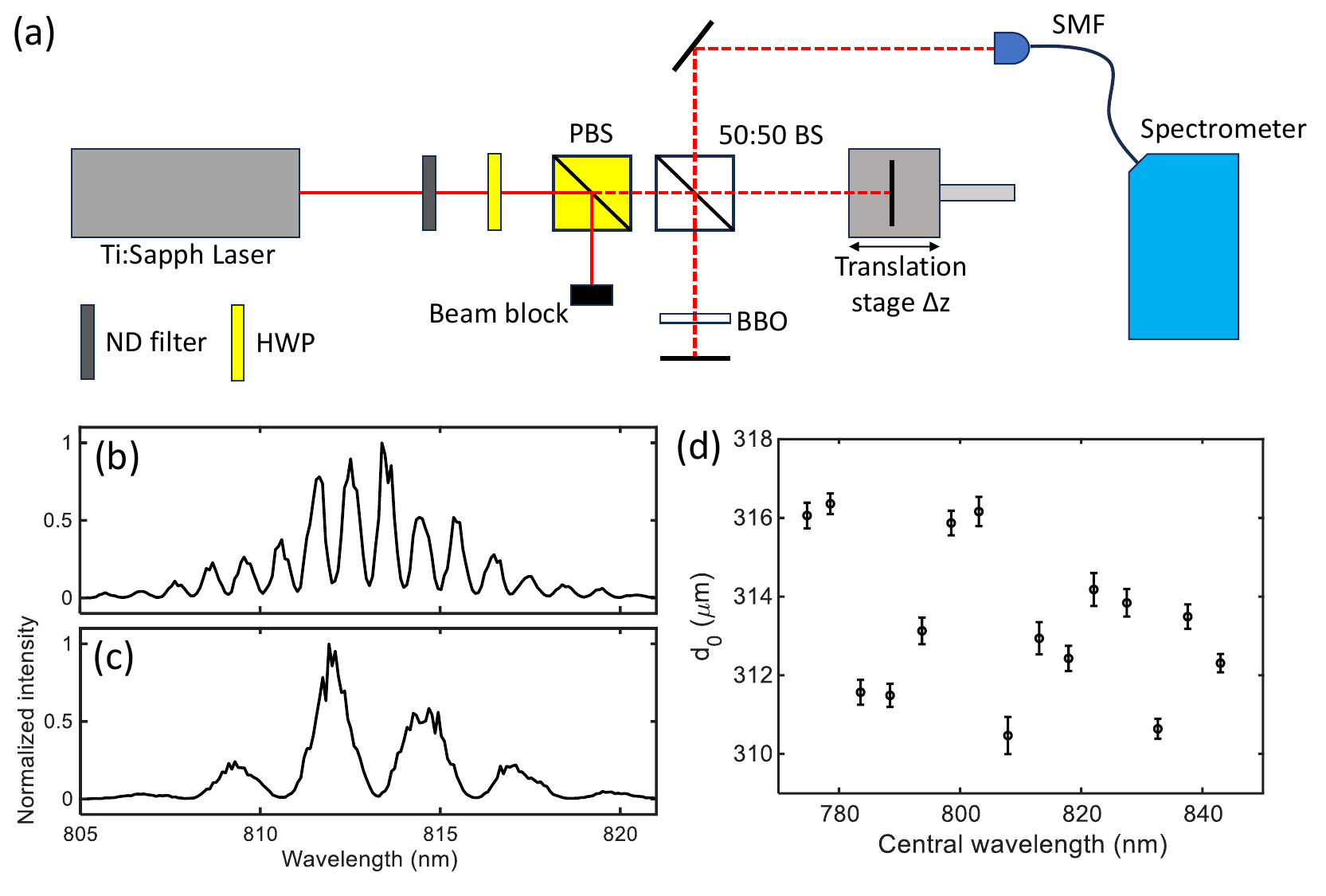}
    \caption{(a) Setup of spectrally resolved interferometry in a Michelson interferometer using a Titanium-sapphire laser. (b,c) Spectrum showing the interference fringe with the BBO crystal (b) and the reference without the crystal (c) at a central wavelength of 813.1\,nm. (d) Measured BBO crystal thickness as a function of the laser central wavelength. }
    \label{Supp5}
\end{figure}

The laser was directed through a Michelson interferometer, with the BBO crystal inserted into one of the interferometer arms, as illustrated in Fig.~\ref{Supp5}(a). In spectral interferometry, the measured interference spectrum can be expressed as
\begin{equation}
    I(\nu)=I_0(\nu)\left[1+V\cos\left(2\pi\nu\Delta t+\phi_0\right)\right],
\end{equation}
where $I_0(\nu)$ is the spectral intensity envelope, $V$ is the interference visibility, $\nu$ is the optical frequency, $\phi_0$ is a constant phase offset, and $\Delta t = \Delta z/c$ is the optical delay difference between the two interferometer arms. The spectral fringe periodicity is therefore determined by the optical path difference, with $\Delta z=c/\Delta\nu$, where $\Delta\nu$ is the fringe period in optical frequency. Thus, performing FFT on the interference spectrum with respect to $\nu$ produces peaks at delays $\tau=\pm\Delta z/c$. The spectra was first converted to uniformly spaced optical frequencies and a Hann window was applied to suppress spectral leakage.

Measurements were performed at multiple laser central wavelengths spanning 774–845\,nm, with an integration time of 50\,ms and 100 repeated measurements at each wavelength. For each central wavelength, a reference measurement was acquired without the BBO crystal, corresponding to an optical path difference of approximately $\Delta z_{\mathrm{ref}}=250$,$\mu$m. The additional optical path difference introduced by the BBO crystal was determined by comparing measurements with and without the crystal. A typical spectrum seen with and without the crystal at a central wavelength of 813.1\,nm is shown in Fig.~\ref{Supp5}(b,c) respectively.

As seen in Fig.~\ref{Supp5}(d), the measured thickness exhibited variations between different central wavelengths that were substantially larger than the statistical uncertainties obtained from repeated measurements at any individual wavelength. These wavelength-dependent fluctuations may arise from systematic effects such as spectrometer wavelength-calibration errors and changes in laser beam pointing during wavelength tuning. However, identifying their precise origin is beyond the scope of the present work.

Averaging the thickness measurements across the investigated wavelength range, using the Sellmeier relation of Eq.~\ref{Sell}, gives $d_0=313\pm2$\,$\mu$m, where the uncertainty represents the observed wavelength-to-wavelength variation in the inferred thickness. 

\section*{Effect of interferometer instability}
In this experiment the interferometer is not actively stabilized, and slow environmental fluctuations such as mechanical vibrations and thermal drift can introduce low-frequency variations in the interferometer path length during long acquisitions. 

\begin{figure}
    \centering
    \includegraphics[width=0.8\linewidth]{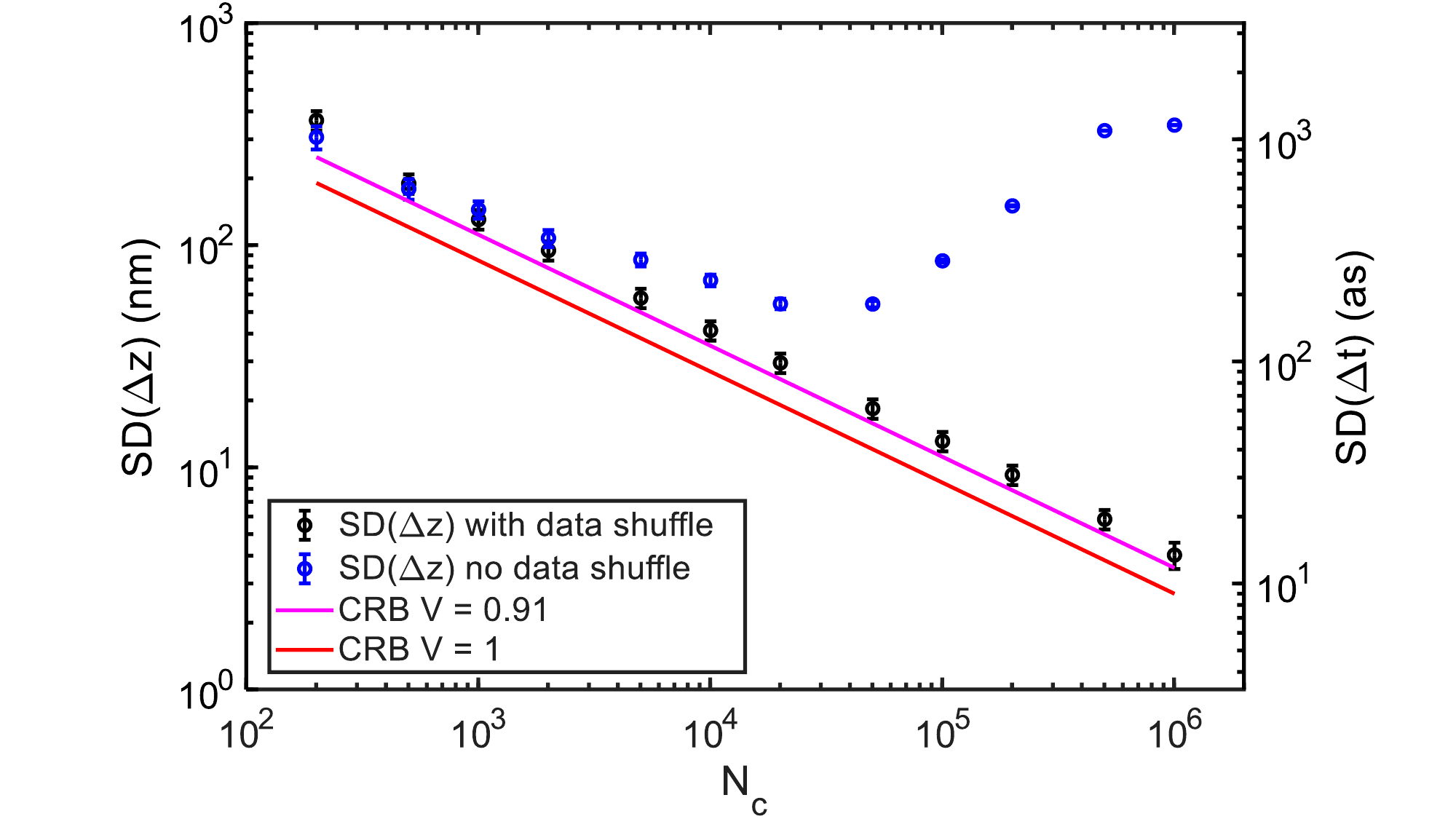}
    \caption{Standard deviation of the measured displacement $\mathrm{SD}(\Delta z)$ as a function of the number of detected coincidences $N_c$, shown for data processed with (black) and without (blue) random shuffling of photon pairs detection order.}
    \label{Supp6}
\end{figure}

To suppress the influence of these fluctuations, we record a long continuous dataset of photon detection events and randomly shuffle the detected photon pairs in time before grouping them into individual measurement batches. This procedure effectively averages over slow interferometer drift, preventing temporal correlations from biasing the delay estimates. The impact of data shuffling is shown in Fig.\,\ref{Supp6}. For small data batches (up to $2000$ coincidence events, corresponding to $\sim 200$\,ms acquisition time), the measured standard deviation in $z$ is identical with and without shuffling, indicating that the measurement remains limited by photon statistics. This transition point also provides an estimate of the dominant environmental fluctuation timescale, corresponding to vibration frequencies on the order of 10\,HZ and lower. However, for larger data batches, the unshuffled data begins to deviate from the expected CRB scaling due to accumulated drift, resulting in a degradation of measurement precision. In contrast, the shuffled data continues to follow the CRB over the full range of $N_c$, demonstrating that the shuffling procedure effectively mitigates the impact of these low-frequency fluctuations.

\end{document}